\documentclass[twocolumn,twocolappendix]{aastex63}
\usepackage[utf8]{inputenc}
\usepackage{placeins}
\usepackage{graphicx}
\usepackage{amsmath}
\usepackage{physics}
\usepackage{mathtools}

\newcommand{\msunpccubed}{M$_\odot$/pc$^{3}$}
\newcommand{\rhodmnospace}{$\rho_{\mathrm{DM},0}$}
\newcommand{\rhodm}{$\rho_{\mathrm{DM},0}$ }

\makeatletter
\DeclareRobustCommand{\HI}{%
  \mbox{H\check@mathfonts\fontsize\sf@size\z@\selectfont I}%
}
\makeatother

\begin{document}

\title{An Acceleration is Worth a Hundred Thousand Phase Space Measurements}

\author[0000-0002-7746-8993]{Thomas Donlon II}
\affiliation{Department of Physics and Astronomy, University of Alabama in Huntsville, 301 North Sparkman Drive, Huntsville, AL 35816, USA}
\affiliation{Minnesota Institute for Astrophysics, University of Minnesota, 116 Church Street SE, Minneapolis, MN 55455, USA}
\affiliation{School of Physics and Astronomy, University of Minnesota Twin Cities, 116 Church Street S.E., Minneapolis, MN 55455, USA}
\correspondingauthor{Thomas Donlon II}
\email{thomas.donlon@uah.edu}

\author[0000-0001-6711-8140]{Sukanya Chakrabarti}
\affiliation{Department of Physics and Astronomy, University of Alabama in Huntsville, 301 North Sparkman Drive, Huntsville, AL 35816, USA}

\author[0000-0003-2676-8344]{Elena D'Onghia}
\affiliation{Department of Astronomy, University of Wisconsin, Madison, WI 53706, USA}

\begin{abstract}
It is now possible to directly measure the accelerations that arise from the distribution of matter in the Milky Way. Direct acceleration studies have already measured fundamental Galactic properties in limited samples such as the local dark matter density.  These acceleration-based measurements of the local dark matter density are now becoming competitive with estimates obtained through traditional kinematic techniques such as Jeans modeling. While classical methods can now draw on the positions and velocities of many millions of stars, recent acceleration-based studies have used fewer than 100 sources, yet achieve comparable precision. A key limitation of kinematic approaches is their reliance on assumptions of dynamical equilibrium and symmetry; direct acceleration measurements do not inherently suffer from this constraint, and are expected to perform better per-source than classical techniques. We find that, for the specific problem of estimating the local dark matter density, a single direct acceleration measurement can provide information comparable to $\sim$10$^5$ stars (and in some cases up to 10$^7$). To explore what leads to this discrepancy, we test the theoretical performance of direct acceleration techniques and Jeans modeling in estimating the local dark matter density using hydrodynamical N-body simulations of a MW-like galaxy, both in isolation and including a Sagittarius-like dwarf to generate disequilibrium. In the equilibrium scenario, Jeans modeling requires one thousand times more sources to achieve the same precision as the direct acceleration approach. This confirms that the per-source information advantage is intrinsic, and does not require disequilibrium to manifest. However, in the perturbed disk, the acceleration-based approach outperforms the Jeans analysis regardless of how many stars are available, due to significant bias in the result inferred from kinematics alone. Our results support earlier findings that non-equilibrium dynamics in the Galactic disk cause Jeans-based methods to systematically overestimate the local dark matter density; we show that this issue may also be present in other types of kinematic studies. \\\vspace{0.5cm}
\end{abstract}

\section{Introduction} \label{sec:intro}

The disk of the Milky Way (MW) contains a wealth of dynamical substructure \citep{Kawata2024}. The present-day structure of the Galaxy arises from the interplay of many different complex physical processes \citep{HuntVasiliev2025}, making its study a valuable probe of fundamental physics in various fields. Among these fundamental properties is the density of dark matter in the Galactic midplane (\rhodmnospace); this is a frequently sought-after measurement in astrophysics because it potentially links the observable behavior of the MW disk to as-yet undiscovered physics \citep{Read2014}, and can inform direct and indirect detection experiments for dark matter \citep{Feng2010,Gaskins2016,TulinYu2018}. 

Estimating \rhodm from the velocities of tracer objects has historically followed the same general procedure: first, one estimates the vertical force at some distance from the midplane, which is subsequently used to estimate the total surface density of the disk. Finally, the measured surface density of stars and gas can be subtracted from this to obtain only the contribution due to dark matter. Variations of this method have been common practice for nearly a century (\citealt{Oort1932}; see \citealt{BertoneHooper2018} for a thorough historical review of the development of this idea). 

While there are many ways to go about estimating the vertical force, the most well-known is Jeans modeling, where one makes use of the Jeans equations to convert between the kinematics of stars at different heights and the corresponding surface density \citep{Oort1932, KuijkenGilmore1991, BovyTremaine2012, Garbari2012, Zhang2013, Buch2019}. In their base form, the Jeans equations arise from combining the collisionless Boltzmann equation with the Poisson equation for gravity, which produces a system of three partial differential equations but nine unknown variables \citep{BinneyTremaine2008}. Because the Jeans equations are not well-posed, in order to obtain meaningful solutions to these equations one must make simplifying assumptions about the distribution function. In Galactic dynamics, this is typically done by presuming that the distribution of stars in the Galaxy is spherically or azimuthally symmetric and is time-independent -- or, in other words, that the disk and dark halo are in dynamical equilibrium. These assumptions allow for the estimation of \rhodm from the vertical kinematics of tracer stars.

Over the last several decades, alternative methods have been developed which use kinematic data of stars to estimate the vertical force and/or surface density, or to otherwise provide an approximation of the distribution function as a whole, without relying directly on the Jeans equations. This produces measurements of \rhodm with independent datasets and different systematics. The exact benefits and drawbacks of each method can vary, although they all suffer from the same limitation; they rely on the present-day observed positions and velocities of stars. 

To understand why this is a restriction, consider a star moving within the MW. The star's orbit can be approximated by integrating the star's position and velocity forward in time. If the distribution function is time-independent, then all past and future kinematic information for this star is encoded within this orbit (plus the presumed distribution function). However, we know that the MW is substantially out of dynamical equilibrium due to mass accretion, perturbations from satellite dwarf galaxies \citep{Donghia2010,Donghia16}, and internal dynamical evolution \citep{Widrow2012,Xu2015,Antoja2018,HuntVasiliev2025,Poggio2025}. As a result, the present-day kinematic information of a given star actually contains imprints of past perturbations, the present potential, and the current distribution function. This means that the supposed orbit of the star according to the present-day potential is not equal to its true historical orbit \citep[this is the concept of ``orbital memory,'' e.g.][]{Arora2022}, and methods which presume time-independence of the distribution function will produce inaccurate answers for the kinematic history of the star. If the distribution function is not known (as is generally the case), using the present-day kinematic information of stars without including the perturbations to the potential over time will lead to an inaccurate recovery of the distribution function. Put simply, a galaxy that is out of equilibrium violates assumptions of symmetry, leading to systematic errors in kinematic studies.  

Given this restriction, an alternative approach is to directly measure the gravitational force at different locations throughout the Galaxy, then use these measurements to infer the distribution of material in the Galaxy via the Poisson equation for gravity. An acceleration contains instantaneous and nonlocal information about the present state of the Galaxy -- unlike kinematic data, which is a convolution of the dynamical history of each tracer object plus the time-averaged potential. The implication of this fact is that direct acceleration measurements allow one to study the distribution function of the galaxy with a temporal and spatial precision that is not possible using kinematic measurements alone. In other words, kinematic methods will only ever measure spatially- and time-averaged properties of the Galaxy. 

Such direct measurements are now possible using the extremely precise time-series observations of dozens of millisecond pulsars \citep{Chakrabarti2021,Donlon2024,Donlon2025}. The acceleration of a pulsar leads to a measurable drift between the expected and observed pulse arrival times over timescales of many years, providing a measurement of the MW's gravitational field at that point. Because these direct acceleration measurements correspond to the gravitational force due to the distribution of matter within the Galaxy in real-time, they do not suffer from the same equilibrium assumptions as kinematic data. Recent theoretical advances allow for non-parametric determination of the local (dark) matter density from pulsar timing data as a function of heliocentric distance, as well as measurements of asymmetries in the dark matter content on either side of the Sun \citep{Donlon2026}. However, the approach of fitting analytical potential models to the observed accelerations still produces the most precise acceleration-based measurements of $\rho_\mathrm{DM,0}$ at this time.  Pulsar timing measurements also now provide a quantitative, but tentative evidence of a dark matter sub-halo near the Sun \citep{Chakrabartietal2026}.  Alternate methods of detecting dark matter sub-halos in the Galaxy using gaps in stellar streams \citep{Bonacaetal2020} lead to mass uncertainties of two orders of magnitude, and there may be other effects in the time-evolving disk that can mimic these features \citep{Arora2026}.

Additional methods for directly measuring accelerations have been proposed -- and surveys are ongoing -- in addition to timing drift of millisecond pulsars. These include extreme-precision radial velocity measurements, where the observed wavelengths of spectral lines for a star change over time due to drifts in the object's line-of-sight velocity due to the Galactic acceleration \citep{Chakrabarti2020}; eclipse timing, which similarly experiences a timing delay in the eclipse midpoint of a binary system due to the Galaxy's gravitational field \citep{Chakrabarti2022}; and double white dwarf binaries, in which an acceleration causes the observed pulsation rate of a white dwarf binary to drift over time, and/or the gravitational waves emitted by the system would experience a secular frequency drift, which will be measurable by LISA \citep{Ebadi2025}. Additionally, it may become possible in the future to obtain tangential (``proper'') accelerations from very precise astrometry of stars \citep{SilverwoodEasther2019}.  At this time, only millisecond pulsars have successfully been used to directly measure an acceleration, although it appears likely that these other methods will become viable within a decade. 

We point out that accelerations have been measured using the shape of tidal streams \citep{Nibauer2025}; however, accelerations inferred from this method are still time- and space-averaged measurements of the accelerations felt across the stream throughout its orbit, rather than an acceleration at a single point at the present day. Studies of stellar streams are able to constrain many non-equilibrium effects (examples include rotation of the bar, dark subhalo interactions, time evolution of the dark matter halo, \citealt{BonacaPrice-Whelan2025}). While undoubtedly useful, because stellar streams are extended objects and we are viewing a single snapshot of their kinematics, these measurements do not exactly correspond to the direct acceleration measurements from individual tracer objects.

It is clear that direct acceleration measurements are valuable, but exactly how they stack up to kinematic measurements has not been determined. Obtaining a single direct acceleration measurement takes a significant amount of time and effort (some pulsars have been observed semi-continuously for decades) compared to determining the kinematic information for a single star (the \textit{Gaia} mission obtained astrometry for billions of stars within a few years). Such a question is crucial for planning future missions that intend to explore the nature of our Galaxy and its dark matter content \citep{Drlica-Wagner2022,Chakrabarti2022whitepaper}. When judging the scientific merit of a project, it would be beneficial to know how much we expect to learn about our Galaxy from a mission which has significant synergy with direct acceleration measurements relative to a more traditional mission that captures kinematic information for many stars. For example, data from high-precision eclipse timing missions (such as \textit{Kepler} or \textit{TESS}) or precise radial velocity observations, for which the primary science driver is detecting exoplanets, can also be used to obtain direct accelerations for stars \citep{SilverwoodEasther2019,Chakrabarti2020,Chakrabarti2022}.

This new context provides projects that are traditionally focused on exoplanet and pulsar timing science with an additional, powerful synergy to "real-time" Galactic dynamics. With this in mind, the primary question for this study is: ``How many stars provide the same amount of dynamical information as a single direct acceleration measurement?'' 

We emphasize that this question is problem-dependent. As it is not feasible to compare every possible use case for these data in a single paper, here we focus specifically on assessing how well kinematic and acceleration data can recover the local dark matter density. This provides an intuition for the dynamical differences between the two types of methods without being overly reductive or restrictive.

The comparison between direct acceleration measurements and kinematic methods rests on two distinct arguments. First, there is an intrinsic information advantage: even in a perfectly equilibrium system, a single acceleration constrains the local gravitational field more directly than any number of stellar phase-space measurements, because it reflects the present-day potential rather than the time-averaged distribution function. Second, there is a robustness advantage: acceleration measurements remain unbiased in a disequilibrium disk where Jeans-based methods systematically fail, because they do not inherit the orbital memory of perturbed tracers. We address both arguments in turn.

In this study, we answer the above question in two ways. First, we use a comparison of local dark matter density measurements from studies over the last decade to assess the relative performance of stellar kinematics and direct acceleration measurements in practice. This accounts for both the variety of methods and tracers used to measure \rhodmnospace, as well as various treatments of the uncertainty in these measurements and reported values. For this observable, we find that across the literature the information gained by adding a single direct acceleration measurement is comparable to the information gained by adding between 10$^3$ and 10$^7$ stars, depending on the dataset and method used. Second, we use $N$-body simulations to directly compute the theoretical performance of the stellar kinematics and direct acceleration measurements. This is done for both an isolated and interacting MW-like galaxy in order to quantify the impact of disequilibrium on each type of measurement. In the isolated disk, which roughly approximates an equilibrium distribution function, a single direct acceleration measurement is worth the same as the full 6-dimensional phase-space information of (at least) one thousand stars when measuring the local dark matter density. In the interacting (disequilibrium) case, direct acceleration measurements dramatically outperform stellar kinematic data at recovering the correct value of \rhodm by as much as 1-2 orders of magnitude, even with a modest number of direct acceleration datapoints.

\section{Literature Analysis}

\begin{figure}[]
    \centering
    \includegraphics[width=0.8\linewidth]{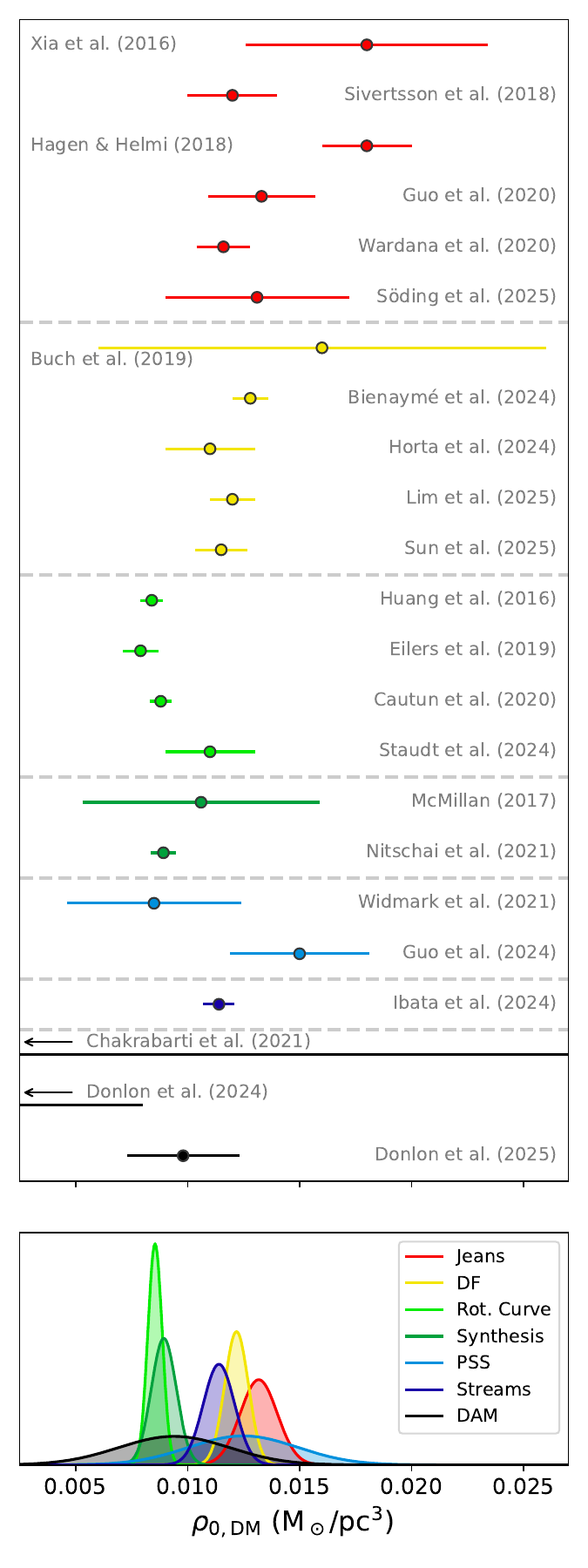}
    \caption{Various literature measurements of the dark matter density in the midplane ($\rho_\mathrm{0,DM}$) from the last decade. Different colors correspond to different types of measurements (see text for details). The bottom panel contains the total joint probability distribution for each type of measurement -- note that some types of measurements consistently measure a higher or lower value for \rhodmnospace. Although early measurements of \rhodm from direct acceleration measurements (DAMs) are located off the plot to the left, the most recent estimate of \rhodm from DAMs is consistent with the other measurement methods.}
    \label{fig:rho_dm}
\end{figure}

The most straightforward way to assess the performance of direct acceleration measurements and various kinematic/dynamical methods at measuring the local density of dark matter is to collate many independent measurements from the last ten years. Each study will independently handle the uncertainties associated with its chosen method and dataset, which provides a more realistic treatment of each problem than we would be able to achieve in a single theoretical study alone.

Figure \ref{fig:rho_dm} shows a comparison of various literature values for $\rho_{0,\mathrm{DM}}$ and the recent values obtained using pulsar direct acceleration measurements (references and data can be found in Appendix \ref{sec:app_fig1_references})\footnote{For a particularly extensive literature review, we direct the reader to the impressive Figure 7 of \cite{Putney2025}. We point out that our general conclusions in this section are still clearly visible in the different types of studies from their plot; our results do not depend on the particular collection of studies we show here.}.  The first direct acceleration results from pulsar timing \citep{Chakrabarti2021} could not constrain $\rho_{0,\mathrm{DM}}$ due to the limited dataset at that time.  Due to the small dynamic range of the available pulsars at that time,  so-called ``$\alpha$ models'' were employed for the potential along with external baryon budgets.  Acceleration measurements become more precise with time, and the updated measurement of \cite{Donlon2025} used an expanded dataset and more realistic MW models, leading to a much smaller uncertainty compared to previous pulsar-based measurements, which is in good agreement with the kinematic measurements of $\rho_{0,\mathrm{DM}}$.

We briefly describe the various categories of kinematic and dynamical methods below: \begin{itemize}
    \item \textit{Jeans modeling}: This has been discussed in detail above, and consists of measuring the velocity dispersion of tracer stars, then using the Jeans equations (along with simplifying assumptions about the distribution function) to convert this to a surface density. Subtracting the amount of stars and gas from this surface density produces the local dark matter density. 
    \item \textit{Distribution function (DF) fitting}: An analytical form is assumed for the distribution function of the positions and velocities of stars throughout the Galaxy, which is then fit to kinematic observations. The local density of dark matter can be obtained as in Jeans modeling, or it can be explicitly fit as part of the distribution function. Stellar metallicity and/or chemical abundances can also be included in the distribution function, a technique known as ``orbital torus imaging'' \citep{Price-Whelan2021}.
    \item \textit{Rotation Curve}: The rotation curve is measured for stars and gas in the Galactic disk. The individual components of the MW (including a dark matter halo) are then fit to this rotation curve plus observations of stars and gas. 
    \item \textit{Synthesis}: Studies which compile various measurements of the surface density of different tracers and/or components of the Galaxy, and then compare these to estimate a value for \rhodmnospace.
    \item \textit{Phase Space Spiral (PSS)}: The phase space spiral is a structure in $z$-$v_z$ space that can arise from a perturbation to an equilibrium disk, which then winds up over time. How much and how quickly this feature winds contains information about the structure of the disk, enabling a measurement of the total surface density and therefore a value of \rhodmnospace.
    \item \textit{Streams}: The orbits of stellar streams are sensitive to the structure of the individual mass components of the MW, including its dark matter halo. While stream data can independently produce measurements of the local dark matter density \citep{Ibata2024}, joint constraints between streams and pulsar accelerations are able to provide improved information on the MW's dark matter content \citep{Craig2023}. Recently, measurement of the Galactic accelerations of individual tidal streams also enable tracing the gravitational field at large distances \citep{Nibauer2025}.
\end{itemize}

\subsection{Systematic Uncertainties}

Different kinematic methods carry systematic uncertainties in their reported measurements of \rhodmnospace. These uncertainties arise from the choice of tracer, statistical uncertainty in the data used, and/or theoretical limitations of the method, and will be different for each study. For example, while Jeans modeling is theoretically exact when used for a time-independent distribution function, it is well-known that Jeans modeling incorrectly estimates the local dark matter density for a disk that has a surface density which varies as a function of time and position \citep{Banik2017,Xiang2018,HainesDonghia2019,SivertssonRead2022}. Additionally, Jeans modeling is understood to produce different values for the surface density when different tracer stars are used \citep{Xiang2018,Cheng2024}. 

Similar drawbacks are present for each method of measuring \rhodmnospace, and each study must make decisions about how to handle these issues in a way that suits their specific needs. This includes direct acceleration measurements -- however, we point out that any systematic uncertainties of direct acceleration studies are entirely independent of systematic issues with kinematic studies, because there is no overlap between their datasets or methods. As a result, the recent pulsar timing studies provide a distinct determination of the local dark matter density. 

Each study in Figure \ref{fig:rho_dm} reports a statistical uncertainty along with their value of \rhodmnospace; by multiplying together the probability distributions of each individual measurement of \rhodmnospace, we arrive at a combined probability distribution for each method. These probability distributions are shown at the bottom of Figure \ref{fig:rho_dm}. 

Interestingly, the kinematic methods for determining \rhodm appear to cluster into two groups, with each group predicting a different, non-overlapping value for the local dark matter density. The first group, consisting of rotation curve and synthesis studies, estimates the \rhodm to be roughly 0.008$-$0.009 \msunpccubed; the second group, consisting of Jeans analysis, distribution function fitting, and stellar stream constraints instead estimate \rhodm to be about 0.012 \msunpccubed. Studies which use the phase space spiral to estimate the local dark matter density have large statistical uncertainties, but the mean value of the combined probability density for these studies agrees with the second group of methods. Similarly, the pulsar studies have large systematic uncertainties in the current data, but are more consistent with the first group of methods. 

The existence of these two groups suggests that one or both groups of methods suffers from a systematic error, or bias, in the reported value of the local dark matter density. We point out that the Jeans estimates of the Oort limit (and therefore the measurement of \rhodmnospace) are expected to be biased high because the over-dense regions of the disk dominate the spatially-averaged analysis \citep{Xiang2018,HainesDonghia2019}. Although we are not aware of any studies that explicitly search for a bias in other kinematic techniques, we speculate that a similar argument might be made for phase space spiral studies, and the same likely holds for distribution function fitting techniques, which will be weighted towards regions with more stars. This suggests that the second group, which claim a higher value for the local dark matter density, may be suffering from systematic bias compared to rotation curve and synthesis studies. 

We point out it is still possible that pulsar, rotation curve, and synthesis studies are biased in some way, although we are unable to identify either the existence of or potential causes for such a bias. As such, it is reasonable to assume at this time that these types of methods do not produce biased measurements of the local dark matter density, unless evidence for a bias is provided in the future.

\subsection{Information per Data-point}

\begin{figure}
    \centering
    \includegraphics[width=\linewidth]{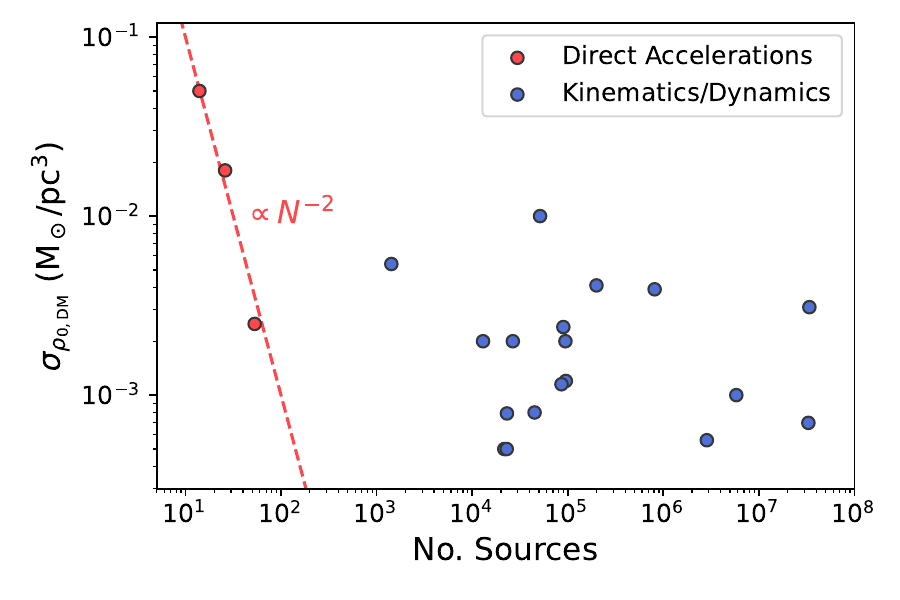}
    \caption{Uncertainty in $\rho_{0,\mathrm{DM}}$ vs. the number of sources required to make that measurement. Data is the same as in Figure \ref{fig:rho_dm}. Red corresponds to direct acceleration (pulsar) measurements, blue corresponds to measurements derived from the kinematics and/or dynamics of stars. While the measurement uncertainty of the kinematics measurements do not obviously decrease with the number of sources used, the uncertainties of the direct acceleration measurements appear to follow a clear power law. If this trend continues, it is expected that direct acceleration studies with as few as 200 sources could provide a $\gtrsim$25$\sigma$ measurement of the local dark matter density, more precise than any existing kinematic measurements (regardless of the available number of stars). }
    \label{fig:unc_scaling}
\end{figure}

\begin{figure}
    \centering
    \includegraphics[width=\linewidth]{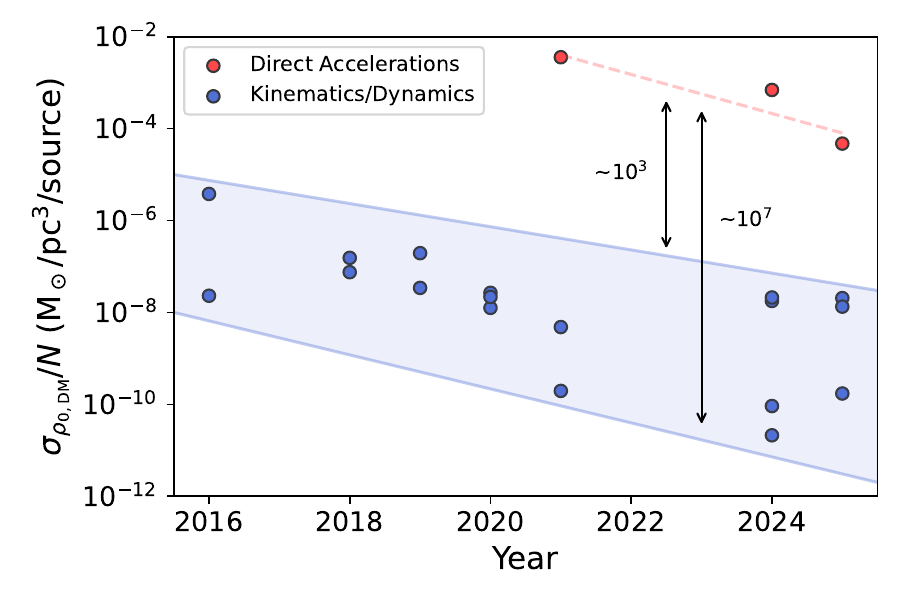}
    \caption{Uncertainty in $\rho_{0,\mathrm{DM}}$ divided by the number of sources used for each measurement (roughly equal to the reduction in uncertainty by adding one new measurement) over time. Colors and data are the same as in Figure \ref{fig:unc_scaling}. The kinematic measurements span a range of values; the reduction in uncertainty from adding a single new direct acceleration measurement is between 3 and 7 orders of magnitude larger than the reduction obtained from adding a single star, depending on the study. }
    \label{fig:contribution_by_year}
\end{figure}

A useful approach for comparing direct acceleration data and kinematic data is to quantify how much information is available in each data-point. While there may be several ways to approach this problem, we use two benchmarks for assessing the information provided by both types of data, specifically in the context of measuring the local dark matter density. 

The first measure of information is to consider the reduction in the uncertainty in an estimate of \rhodm when additional data-points are added. In other words, if the number of pulsar accelerations or stars in a dataset increases, how much more precisely can \rhodm be measured? This is shown for our collection of studies in Figure \ref{fig:unc_scaling}, where the behavior of the kinematic/dynamical studies and the direct acceleration measurement studies are clearly different. In this case, ``information per data-point'' takes the form $I = -\dv*{\sigma_{\rho_{DM,0}}}{N}$. A steeper decrease in the uncertainty of the DM density as one adds more points indicates that each individual data-point contains more information.

We find that the precision of kinematic/dynamical studies does not depend strongly on the number of sources included in the dataset ($N$). This is surprising, since naively one would expect the precision of a measurement to roughly improve as $\sqrt{N}$ according to Poisson statistics. The substantial scatter in the uncertainty for a given number of kinematic sources is likely due to differences in the datasets and methods used by these various studies, which could obscure trends in individual methods. Additionally, if the primary driver of the uncertainty in these studies is systematic uncertainty rather than per-datapoint noise, then improving the number of sources will not have a large effect on the overall uncertainty. 

On the other hand, the pulsar studies show a clear improvement as the number of available sources increases roughly following $N^{-2.0}$, improving much more quickly than expected according to naive Poissonian statistics. If this observed trend continues, then direct acceleration studies would provide more precise measurements of the local dark matter density than kinematic studies with only 200 sources.

This odd ``super-Poissonian'' trend can be explained because the apparent trends for pulsar studies are a combination of both the statistical improvement due to the increase in the number of datapoints, as well as improvements in methodology and reductions in the uncertainties in the positions and accelerations of individual pulsars. Specifically, a general scaling relation for the uncertainty in the measurement of \rhodm can be split into two individual power laws of the form $\sigma_{\rho_\mathrm{DM,0}} \propto N^{-a} T^{-b}$, where $T$ is the baseline observation time of the pulsar data. It must be the case that $a = 1/2$ following Poissonian statistics (this is shown later on in Section \ref{sec:precision_per_source}). The general form of $b$ is not easily calculable, because each pulsar has its own individual scaling relation that depend on that pulsar's position in the Galaxy and on the sky. The relations for the components of each pulsar's acceleration and position are known, however: $\sigma_{\dot{P}_b} \propto T^{-5/2}$, the relative uncertainty in proper motion scales as $T{-3/2}$, and the relative uncertainty in parallax scales as $T^{-1/2}$ \citep{DamourTaylor1992,LorimerKramer2012Handbook}. The final component in $b$ is improvement in methodology, which is difficult to quantify numerically. Appendix B of \cite{Donlon2024} computes a prediction for the improvement in \rhodm given reasonable assumptions for the improvements in individual pulsar properties over time, and predict a scaling relation of roughly $T^{-1.0}$.

Note that when we plot a scaling relationship as a function of only one variable (as in Figure \ref{fig:unc_scaling}), we are condensing the actual scaling behavior of $\sigma_{\rho_\mathrm{DM,0}} \propto N^{-a} T^{-b}$ into something that looks like it is only a function of $N$ or $T$ individually, which can lead to trends that \textit{appear} better than would be statistically possible. In other words, $N^{-1/2}\cdot T(N) \sim N^{-2.0}$, and $N(T)\cdot T^{-b}\sim T^{-1.0}$. This is why we see an apparent trend of $N^{-2.0}$ for the pulsar studies in Figure \ref{fig:unc_scaling}, even though it can be mathematically shown that $\sigma_{\rho_\mathrm{DM,0}}$ must obey a $\sqrt{N}$ scaling relation. These relations are still useful, however; improvement in the number and precision of pulsars is expected to carry on at the current pace (or better), and because the number of available pulsar accelerations is based on how many pulsars we have precise timing solutions for (which improve as the observational baseline $T$ increases), $N$ and $T$ are tied together in a complicated but straightforward way.

The second measure of information is to consider the uncertainty in \rhodm divided by the number of data-points used to obtain that value -- in essence, this is the precision contributed by each data-point in a sample. In this case, ``information per data-point'' takes the form $I = \sigma_{\rho_{DM,0}}/N$. This is shown as a function of time in Figure \ref{fig:contribution_by_year}, and there is a clear vertical separation between the kinematic/dynamical methods and the direct acceleration studies. 

There is scatter in the kinematic/dynamical studies due to the variety of methodology and data used, but overall the precision per source has decreased over time slightly in both the kinematic/dynamical studies and the pulsar studies. However, the direct acceleration measurements have between 10$^3$ and 10$^7$ times more precision per source than the kinematic/dynamical studies (depending on the individual study). We can conclude from this that a single direct acceleration measurement is worth the same amount of information as the phase-space data for at least 10$^3$ stars, although this could be as large as 10$^7$ stars in some cases. 

This makes intuitive sense; for each acceleration datapoint, one would equivalently need to obtain phase space information for many nearby stars in order to estimate the corresponding single acceleration at that point in space using Jeans modeling or a similar method. The simulations in the following sections allow us to separate how much of this per-source advantage is intrinsic to the two measurement types, and how much is amplified by disequilibrium effects that systematically bias kinematic methods.

Despite the gap in the amount of information per datapoint, the change over time in the precision per source are roughly the same for the two types of studies. The per-datapoint precision of kinematic/dynamical studies decreases between roughly $I \propto T^{-0.58}$ and $I \propto T^{-0.85}$ (shown as the bounding blue lines in Figure \ref{fig:contribution_by_year})  while the pulsar studies decrease as approximately $I\propto T^{-0.97} \sim 1/T$ (shown as the dashed red line), in good agreement with the estimates in Appendix B of \cite{Donlon2024}.

\subsection{Number of Sources vs. Source Characterization}

By analyzing Figures \ref{fig:unc_scaling} and \ref{fig:contribution_by_year}, we can begin to distinguish what portion of the improvement in $\sigma_{\rho_{DM,0}}$ is due to increasing the number of data-points vs. improved characterization of individual data-points over time. Figure \ref{fig:unc_scaling} shows that the acceleration data scales strongly with the number of data-points in the sample, but there is no corresponding trend in the kinematic data. However, in Figure \ref{fig:contribution_by_year} the results from kinematic data continue to improve over time (albeit with significant scatter).



This informs us that improvement in determining $\rho_{DM,0}$ primarily (but not solely) comes from increasing the number of data-points for direct acceleration studies at this time. On the other hand, improvement for kinematic studies appears to essentially only come from improvements in the uncertainties on individual data-points, new techniques, and/or better data characterization that is not strongly connected to the number of available data-points. To determine how much of the per-source advantage is intrinsic and how much is driven by the systematic bias that disequilibrium introduces into kinematic methods, we turn to a controlled set of $N$-body simulations.

\section{Simulations}

\begin{figure*}
    \centering
    \includegraphics[width=\linewidth]{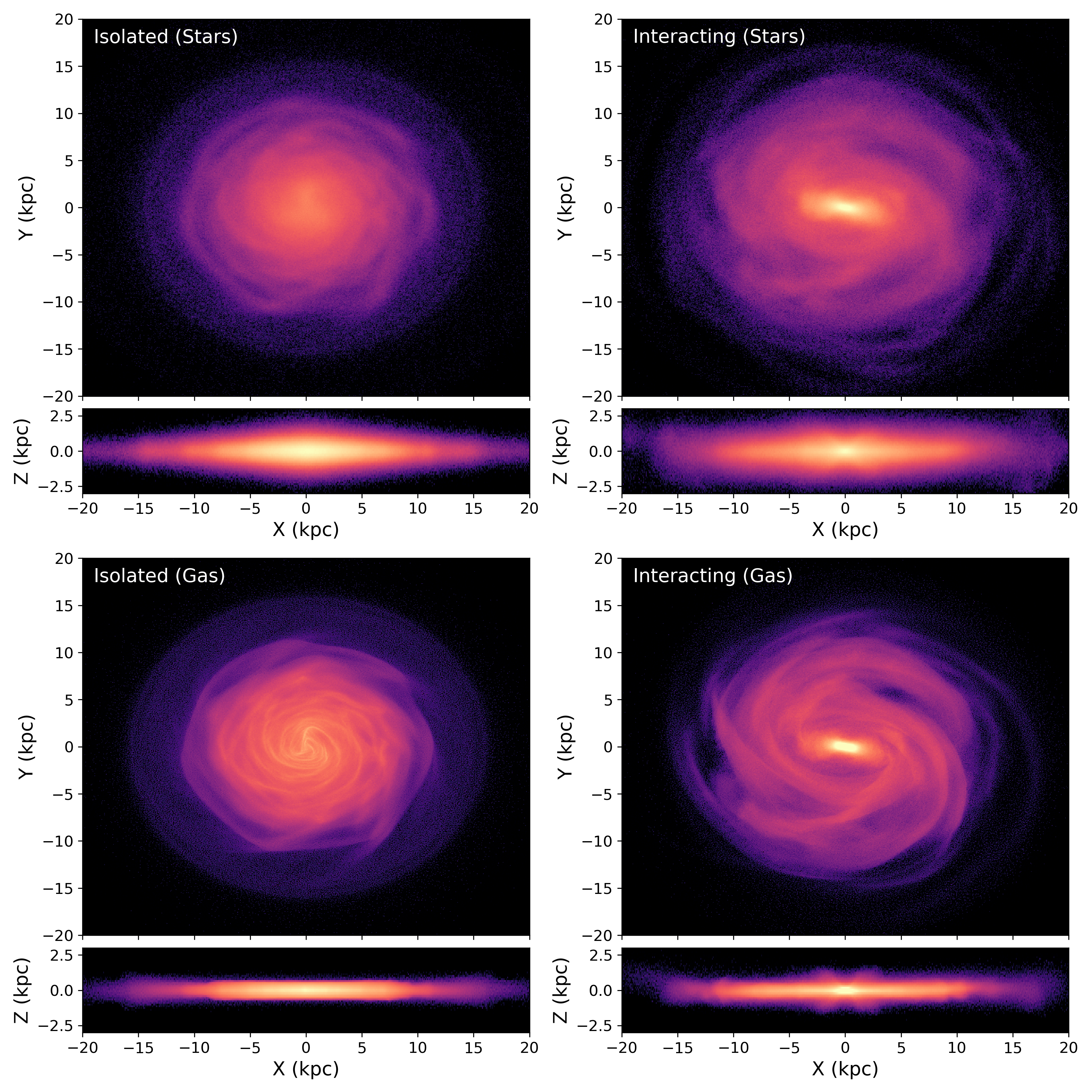}
    \caption{Baryon density of the two simulated galaxies used in this work. The left column shows the isolated galaxy, while the right column shows the same galaxy interacting with a sizable orbiting satellite dwarf galaxy. The top row shows the star particles, and the bottom row shows the gas particles for the two simulations. Face-on and edge-on views are shown for each panel. The isolated galaxy has a small amount of structure, mostly limited to the outer portion of the disk. The interacting galaxy forms a significant (boxy/peanut) bar/bulge and prominent spiral arms that are driving disequilibrium processes in the kinematics of the disk stars.}
    \label{fig:edgeon_faceon}
\end{figure*}

\begin{figure}
    \centering
    \includegraphics[width=\linewidth]{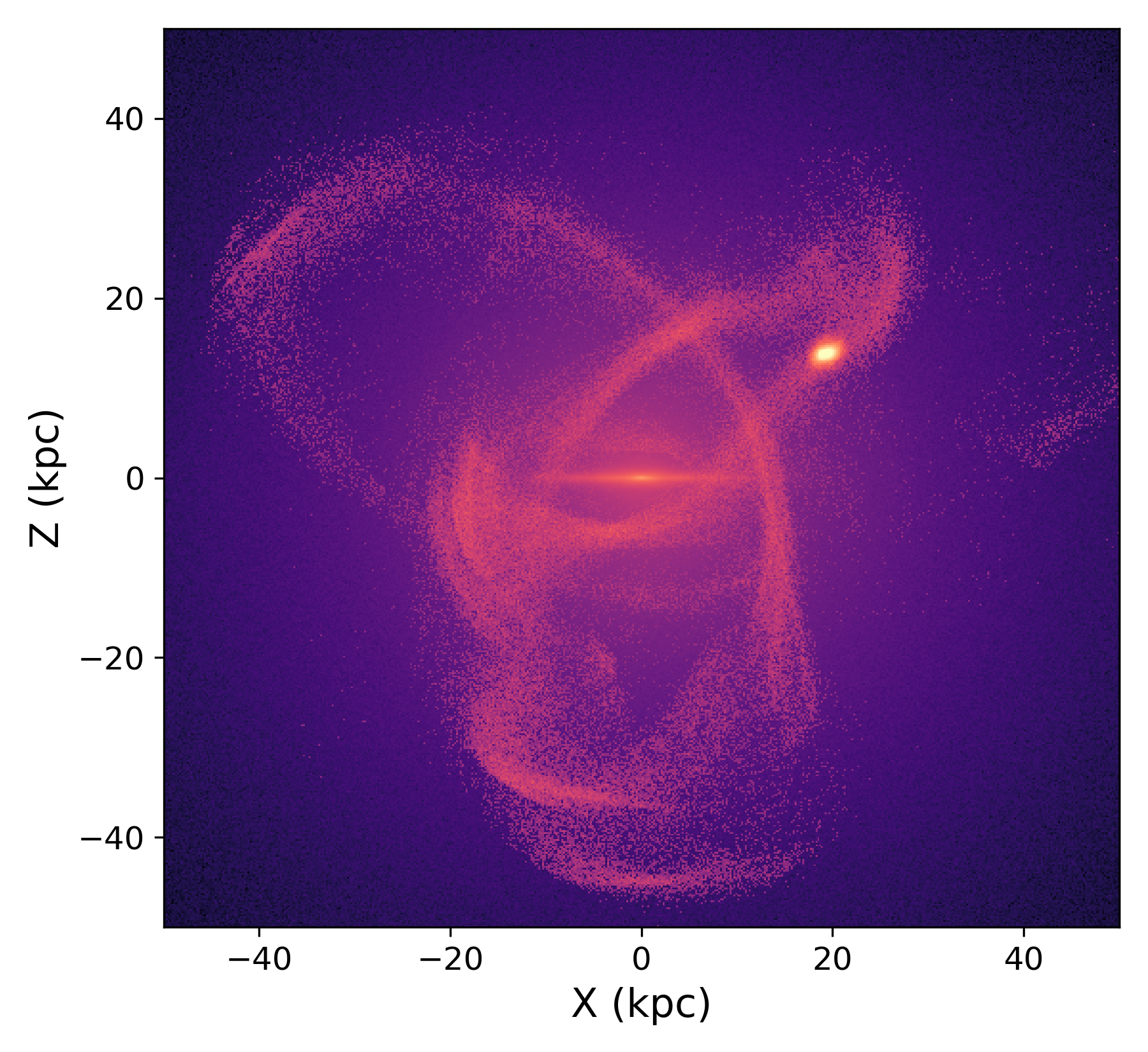}
    \caption{Density of the host galaxy plus the orbiting satellite in the interacting simulation. In order to improve visibility, the density of star particles that initially belonged to the satellite galaxy has been enhanced by a factor of 1000 relative to the host galaxy. The simulated debris vaguely resembles the Sgr stream in the MW due to its trefoil shape and polar orientation. Note that we are not trying to recreate the Sgr stream in this simulation; we are simply trying to perturb the host galaxy disk in a realistic way. }
    \label{fig:satellite}
\end{figure}

\begin{figure*}
    \centering
    \includegraphics[width=\linewidth]{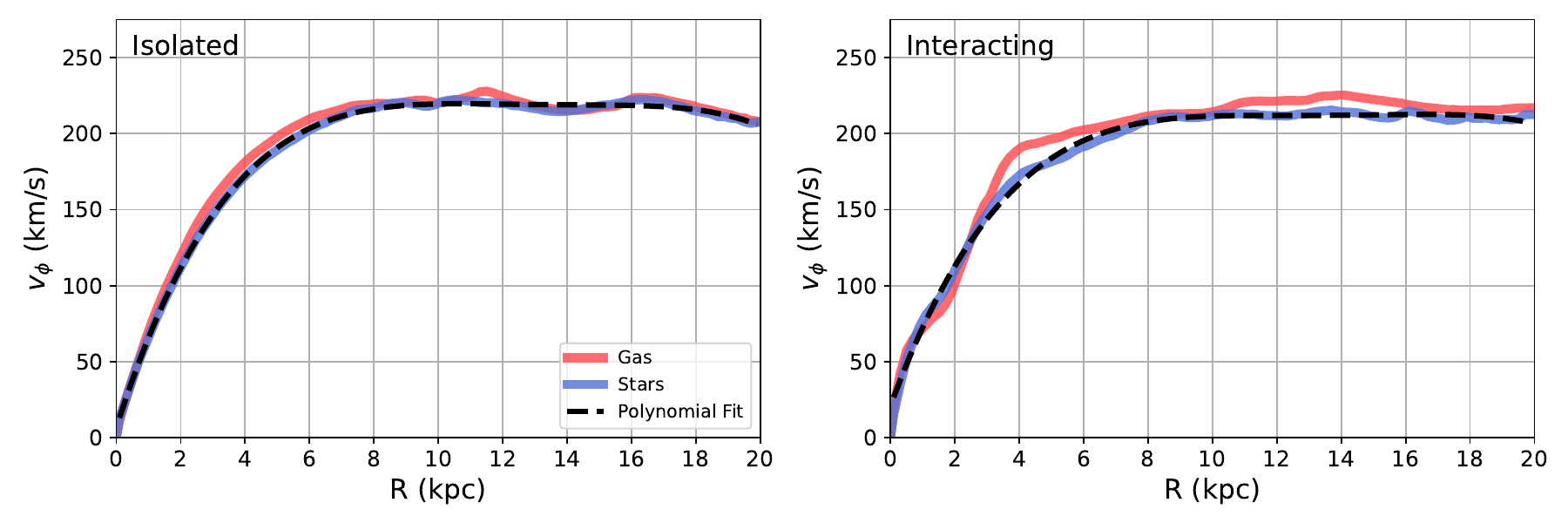}
    \caption{Rotation curves of the two simulations, split up into the star and gas particles (blue and red, respectively). A 4th degree polynomial fit to the stellar data is shown as a dashed black line. In the isolated simulation, the rotation speed of the gas and stars broadly agree at all radii; in the interacting simulation, the gas orbit somewhat more slowly than the stars in the region dominated by the bar ($r<3$ kpc), while the stars orbit somewhat more slowly than the gas outside of this radius.}
    \label{fig:rotcurve}
\end{figure*}

In the second half of this paper, we use hydrodynamical $N$-body simulations of a MW-like galaxy in order to assess the theoretical performance of direct acceleration measurements and Jeans modeling at accurately measuring the local dark matter density. For the purposes of our study, we broadly consider the isolated simulation to approximately be in dynamical equilibrium, while the interacting simulation contains many strong non-equilibrium features, discussed in detail below. 

We run two simulations: a MW-like disk galaxy in isolation, and the same galaxy but with an orbiting satellite galaxy. The simulations were created using a custom version of the GALIC initial condition generator \citep{YurinSpringel2014} with 5 million gas particles, 5 million star particles, and 40 million dark matter particles, for a total number of 50 million particles. The profile of the simulated galaxy is taken to match the host galaxy in the ``100R15'' model of \cite{ChakrabartiBlitz2009, Chakrabarti_Blitz2011}, which was designed to be similar to the MW. 

The stellar component of the simulated galaxy consists of a double exponential stellar disk with an initial scale length of 4.1 kpc and scale height of 400 pc. There is a gas disk with the same vertical and horizontal distribution as the stars and a total mass equal to one-fourth of the stellar disk mass, plus a flat extended \HI\;disk with a scale length of 12.3 kpc and one-half of the mass of the inner gas component. The dark matter halo consists of a Hernquist profile \citep{Hernquist1990} with a concentration of 9.4 and a circular velocity at $r_{200}$ of 160 km/s. The total stellar, gas, and dark matter masses of the host galaxy model are 3.5$\times10^{10}$ M$_{\odot}$, 1.25$\times10^{10}$ M$_{\odot}$, and 1.44$\times10^{12}$ M$_{\odot}$, respectively. 

The interacting simulation includes an orbiting satellite made up of 1 million dark matter particles. The satellite has a mass of $10^{9}$, and is initially generated as an NFW profile \citep{NFW} with a scale radius of 3 kpc, generally consistent with a large spheroidal dwarf galaxy. 

These simulations were run forward in time for 4 Gyr using the Gadget-4 code \citep{Gadget4}. The exact size of each timestep is chosen by Gadget-4 to ensure numerical stability of the particle integration, but the maximum allowed timestep size is 1 Myr. The smooth-particle hydrodynamics follows the pressure-entropy formulation of \cite{Hopkins2013}, and additional star particles are created when gas exceeds temperature and density thresholds according to the default treatment of gas dynamics within Gadget-4 \citep[see][]{SpringelHernquist2003,Gadget4}. The softening parameter of each particle type is allowed to vary throughout the simulation based on the density of particles, but the maximum allowed softening lengths for stars, gas, and dark matter are 25, 50, and 25 pc, respectively. 

The host galaxy was initially generated in dynamical equilibrium in both cases, although relaxation occurs throughout the $N$-body integration that somewhat changes the distribution function even for the isolated galaxy. Namely, the stellar disk becomes thicker, with a scale height of 500-700 pc; the inner gas disk becomes more condensed, with a scale height of about 50-100 pc, and the flat extended \HI\; gas disk develops vertical flaring at large $R$. 

The stellar and gas distributions of these galaxies are shown in Figure \ref{fig:edgeon_faceon}. Although the isolated and interacting simulation have similar scale lengths and heights for the stellar and gas disks, there are some notable differences between the two simulations. In particular, the interacting simulation forms a strong boxy/peanut bulge/bar (which buckles early in the simulation) that is not present in the isolated galaxy. Similarly, the interacting simulation forms broad spiral structure that is also not present in the isolated simulation. The isolated simulation is approximately in dynamical equilibrium, although there are still small variations in the distribution of stars and gas throughout the disk. The interacting galaxy is -- as expected -- clearly not in dynamical equilibrium as it shows strong non-axisymmetric structure, including spiral arms, a large bar, and vertical warping of the disk. 

The orbit of the satellite in the interacting simulation was chosen to be similar to that of the Sgr dwarf galaxy in the MW, because the Sgr dSph is believed to substantially contribute to the observed dynamical disequilibrium features in the MW disk \citep{HuntVasiliev2025}. In order to achieve this, we began with the observed present-day position and velocity of the Sgr dSph in the MW \citep{Vasiliev2021} and integrated a test particle with those phase-space coordinates backward in time for 3 Gyr in the frozen potential of the isolated galaxy simulation; this allows the simulated satellite to ``over-shoot'' the actual Sgr dSph location, so that the majority of the satellite's particles were not located near the disk during our analysis, which could potentially influence our results. The satellite was then generated with the final position and velocity of this test particle's orbit. The simulated orbit of the satellite does not exactly match that of the test particle, because the simulated satellite experiences dynamical friction along its orbit \citep{Chandrasekhar1943}, induces reflex motion in the host galaxy \citep{PetersenPenarrubia2021}, and the live potential of the host galaxy will differ from the frozen isolated potential \citep{Arora2022}. Additionally, the potential of the simulated galaxy is not identical to the potential of the MW, leading to differences in the final stream morphology. We also do not enforce that the position and velocity of the satellite match those of the actual Sgr dSph in the analyzed snapshot. The resulting orbit of the satellite, shown in Figure \ref{fig:satellite}, still vaguely resembles that of the actual Sgr dSph. This is satisfactory for this study, as our goal is not to recreate the exact properties of the Sgr tidal stream, but to induce disequilibrium features in the disk of the simulated host galaxy which are comparable to those in the MW.

The rotation curves of the gas and stars in each simulation are shown in Figure \ref{fig:rotcurve}. In the isolated simulation, the gas and stellar disks have nearly identical rotation curves, while in the interacting simulation the gas disk has a slightly higher rotational velocity than the stars. Both simulations have small dynamical features present in both the stellar and gas rotation curves, although these are more pronounced in the interacting simulation. The overall shape and magnitude of the rotation curve is similar to that of the MW for both simulations \citep{Sofue2020}.

\section{Applying Kinematic and Acceleration Methods to Simulations}

In order to compare direct acceleration methods with kinematic methods, we use both methods to estimate the local density of dark matter in different places throughout the disk in both simulations. We explain below the exact methodology for obtaining a value of \rhodm for both cases. For kinematic data, we only consider Jeans modeling as a fiducial, representative kinematic method for this study; however, note that there are other kinematic methods that may perform better or worse in different scenarios. As shown above, each kinematic method will have its own benefits, drawbacks, and varying levels of accuracy.

\subsection{Jeans Modeling} \label{sec:eq_jeans}

Our Jeans modeling approach is based on the procedure of \cite{HagenHelmi2018}. By combining the Poisson equation with the Jeans equations in cylindrical coordinates, one obtains an estimate for the surface density $\Sigma$: \begin{align} \label{eq:jeans}
    2\pi G \Sigma(R,z) &\simeq \frac{\sigma^2_z}{h_z} - \pdv{\sigma_z^2}{z} \\ \nonumber
    &- \mathrm{cov}(v_R,v_z)\left[\frac{1}{R} - \frac{2}{h_R}\right] + 2 v_c \frac{|z|}{R}\pdv{v_c}{R},
\end{align} where $h_z$ is a kind of ``local scale height'' \citep[see][]{HainesDonghia2019} given by \begin{equation}
    h_z(z) = -\left(\dv{\ln\rho(z)}{z}\right)^{-1}, 
\end{equation} which must be computed individually for a given location on the disk. Note that this procedure makes a few implicit assumptions about the distribution function, namely: (i) it is time-independent, (ii) it is axisymmetric, and (iii) it is symmetric above and below the midplane. These assumptions are expected to break down in the interacting case considered below. 

The last term in Equation \ref{eq:jeans} is a modification added by \cite{HainesDonghia2019}, which corrects for the change in circular speed as a function of height and radius. We point out that the magnitude of this correction is overestimated if one fits the local shape of the rotation curve including any small-scale kinematic features in the nearby disk (see Appendix \ref{app:jeans_error_rotcurve}). This is because the kinematics of disk stars are set by the mean orbital velocity at that star's guiding radius rather than the ``instantaneous'' rotation curve at that star's present location -- the impulse of such a feature on any given star is small, and any minor random fluctuations in structure will average out over the course of the star's orbit around the Galaxy. This issue is resolved by instead considering a smooth approximation to the rotation curve that does not have dips or other noise; in our case, we use a 4th degree polynomial fit to the azimuthally-averaged stellar rotation curve of each simulation rather than the instantaneous rotation curve at any point in the disk. This is shown as a black dashed line in each panel of Figure \ref{fig:rotcurve}.

The total surface density ($\Sigma_\mathrm{tot}$) is calculated by considering stars within a small face-on area of the disk, but all heights. We create a histogram of the $|z|$ positions of these stars (making sure to center the data first such that the mean value of $z$ lies in the midplane), and then compute $h_z$ from the mean slope of the log counts in each bin. The velocity dispersion of the stars is computed independently for each bin, while the derivative of the velocity dispersion is obtained using a best fit line for the velocity dispersion of all bins \citep[following the procedure of][]{HagenHelmi2018}. These approximations are used in place of the true derivatives, which introduce a large amount of noise into the calculation because they must be calculated numerically for a relatively small number of stars.

\begin{figure*}
    \centering
    \includegraphics[width=\linewidth]{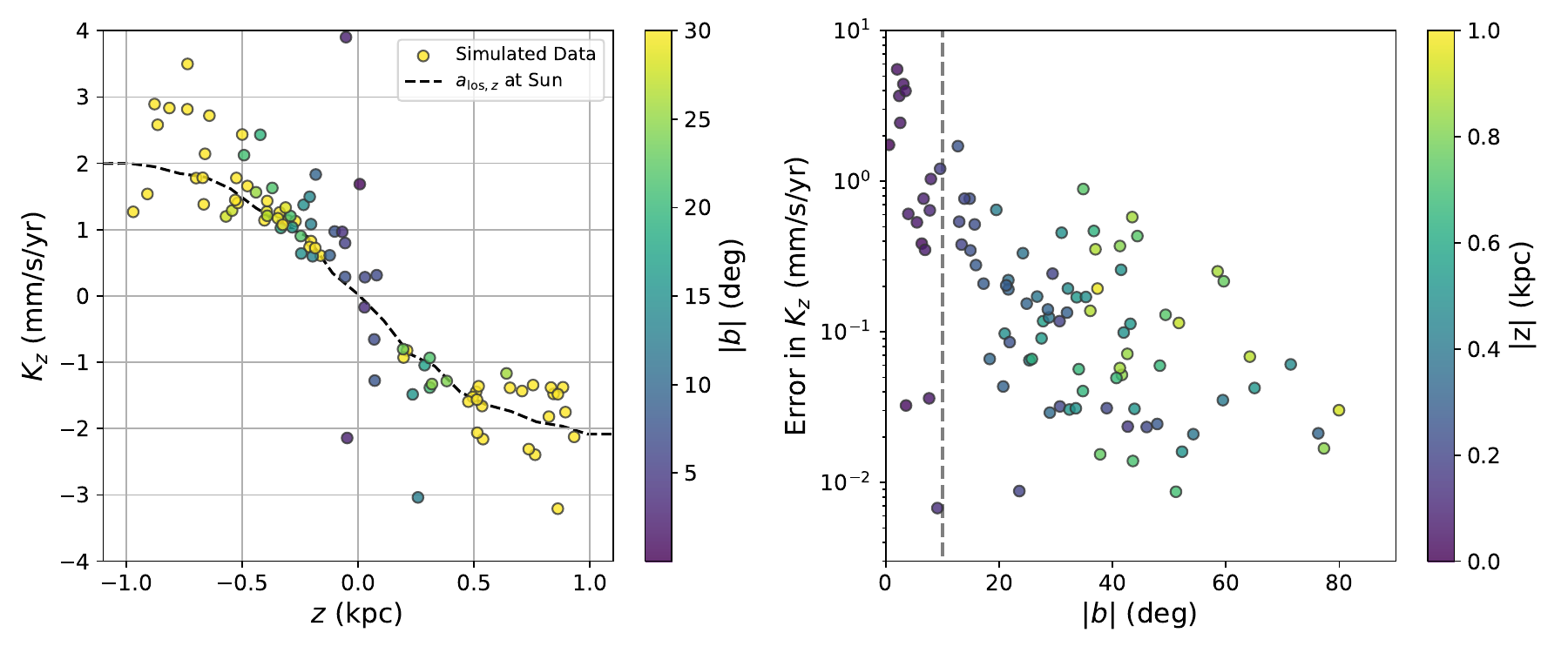}
    \caption{\textit{Left:} Vertical force curve for the isolated simulation (dashed gray line), along with 100 mock accelerations sampled uniformly within a 1 kpc box centered on the Sun. These mock accelerations do not lie exactly on the true line-of-sight acceleration curve due to the projection of the vertical coordinate of the acceleration, and differences in the vertical force curve at different Galactocentric radii. The data are colored according to the Galactic latitude of each point. Points with small values of $|b|$ have large scatter compared to the points with large $|b|$ values. \textit{Right:} Uncertainty in the vertical force due to projection as a function of Galactic latitude. Points are colored according to their distance from the midplane. Data very close to the midplane (small $|b|$) has large uncertainty; we remove points with $|b|<10^\circ$ for this reason (vertical dashed grey line).}
    \label{fig:kz_accel}
\end{figure*}

The circular velocity and its derivative are taken from the dashed black line in Figure \ref{fig:rotcurve}. The scale length of the disk ($h_R$) is computed once for the entire disk in each simulation by fitting a surface density model with exponential fall-off to the radial density of disk stars given by \begin{equation}
    \rho(R) = \int_0^{2\pi}\int_{-\infty}^\infty \rho(R,z,\phi) \; \dd z \; \dd \phi.
\end{equation}

Then, the mean dark matter density within a distance of $|z|$ from the midplane is estimated as \begin{equation} \label{eq:rhodm_from_sigma}
    \rho_{0,DM} \simeq \frac{\Sigma_\mathrm{tot}(z) - \Sigma_*(z) - \Sigma_\mathrm{gas}}{2z},
\end{equation} where we have assumed here that the surface density of stars ($\Sigma_*(z)$) and gas ($\Sigma_\mathrm{gas}$) are known precisely. This is true in the simulation, but in the actual MW these quantities must be measured, and are only known up to some level of precision \citep{McKee2015,BlandHawthornGerhard2016,Bovy2017b,Cautun2020}. This means that Jeans modeling techniques in real data will have additional systematic uncertainties arising from errors in these values, which are not explored here. 

Here we assume that the gas disk is infinitely thin, so that the surface density of the gas is not a function of $z$. This is acceptable in our case, since we primarily compare the mean dark matter density inferred within $|z|=1$ kpc for this study, and the scale height of the gas disk is much smaller than 1 kpc.

\subsection{Direct Acceleration Modeling} \label{sec:direct_acc_modeling}

Here we develop a method to obtain the vertical force (and therefore the surface density) throughout the disk from a collection of direct acceleration measurements. We begin by noting that each source provides a single measurement of the line-of-sight acceleration at the location of that object: \begin{equation}
    a_\mathrm{los}(\mathbf{x}) = \left[\mathbf{a}(\mathbf{x}) - \mathbf{a}(\mathbf{x}_\odot)\right]\cdot \hat{d},
\end{equation} where $\hat{d}$ is the unit line-of-sight vector from the Sun to that source. Note that the quantity being measured is not an absolute acceleration with respect to the Galactic inertial frame, but a relative acceleration with respect to the Sun. 

Writing the above expression out explicitly in terms of the individual acceleration components, we have \begin{equation}
    a_\mathrm{los} = \left[(a_R - a_{R,\odot}) \hat{R} + a_z \hat{z}\right]\cdot \hat{d},
\end{equation} where we have assumed axisymmetry ($a_\phi \rightarrow 0$) for an appropriate comparison to the Jeans modeling, and that the Sun is located in (or very close to) the midplane ($a_{z,\odot} \rightarrow 0$). Since we are interested in the vertical structure of the disk, we solve for $a_z$: \begin{equation} \label{eq:deproj_az}
    a'_z = \frac{1}{|\hat{d}_z|}\left[ a_\mathrm{los} - (a_R - a_{R,\odot})|\hat{d}_R|\right],
\end{equation} where $|\hat{d}_i|$ is the magnitude of the $i$-component of the unit line-of-sight vector, and the radial acceleration is given by \begin{equation}
    \mathbf{a}(R,z=0) = -\frac{v_\mathrm{circ}^2}{R}\hat{R}.
\end{equation} 

\begin{figure}
    \centering
    \includegraphics[width=\linewidth]{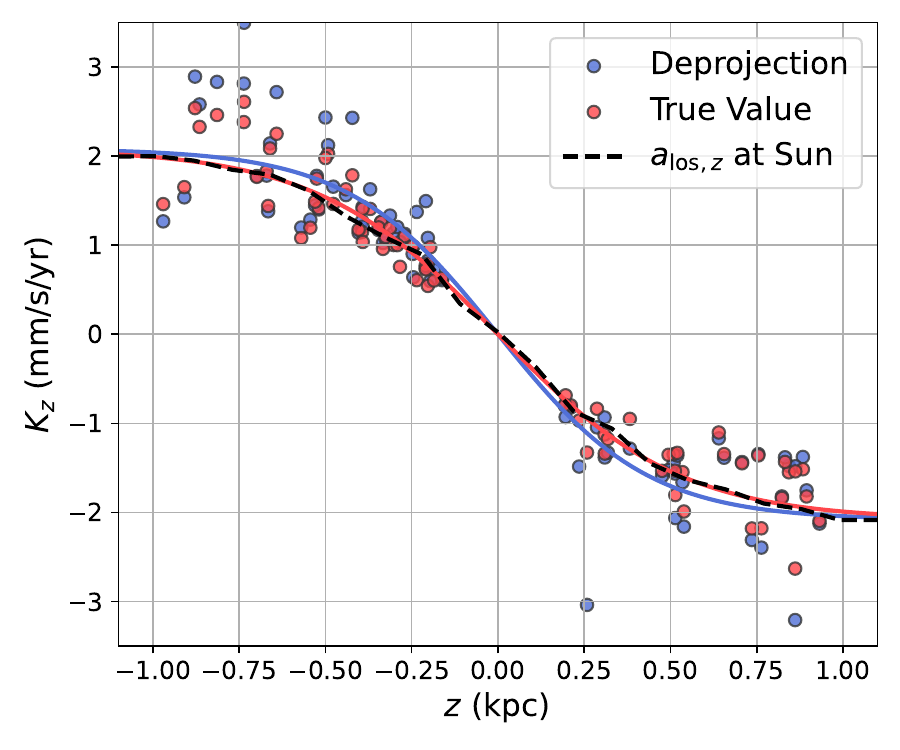}
    \caption{Simulated vertical force data (identical to Figure \ref{fig:kz_accel}). Points with Galactic latitude $|b|<10^\circ$ have been removed. Red points show the true vertical force at each point, and blue points show the estimate for the acceleration obtained by deprojecting the line-of-sight acceleration. Although there is some increased scatter in the deprojection estimates for the vertical force, the vertical force profiles fit to the data (solid lines) are very similar for both the red and blue points. The similarity of the fit to the deprojected data (blue) and the true underlying vertical force profile (black dashed line) shows that we are able to obtain a reasonable estimate for the vertical force curve in this way.}
    \label{fig:proj_vs_true}
\end{figure}

We refer to the quantity on the left-hand side of Equation \ref{eq:deproj_az} as the ``de-projected vertical force,'' and it is an estimate of the true vertical force given the position of the source and its measured line-of-sight acceleration. Since it is an estimate rather than the true vertical acceleration of the source, we label it with a prime to avoid confusion. Mock observations of $a'_z$ from the isolated simulation are shown in Figure \ref{fig:kz_accel}. Note that the de-projected $a_z$ does not exactly agree with the true vertical acceleration of a given source due to projection effects and neglected components, as shown in Figure \ref{fig:proj_vs_true}. In general, the de-projected $a_z$ has a larger magnitude than the true value of $a_z$, leading to a small overestimation of the vertical force. This is partially due to variations in the radial and azimuthal accelerations of each source, which are present in the simulations but not accounted for in our model. Additionally, some sources may be very close to the Galactic plane, which leads to numerical noise when de-projecting the vertical acceleration. For this reason, we only use sources with an apparent latitude of $|b|>10^\circ$ for this procedure. Including all sources results in a mean $K_z$ error of roughly 0.1 mm/s/yr across each sample; removing sources within 10$^\circ$ of the Galactic midplane reduces this to 0.065 mm/s/yr. 

We can then use measurements of $a'_z$ from many sources to estimate the vertical force function, $K_z(z)$, by fitting a parametric model to the simulated data. In this case, because the vertical density distributions of the simulations are roughly $\sech^2$ disks, our model is \begin{equation}
    K_z(z) = -A h_z \tanh\left(\frac{z}{h_z}\right),
\end{equation} where the overall amplitude $A$ and the vertical scale height $h_z$ are free parameters. The local dark matter density is then estimated in the usual way from the value of $K_z(z=1)$ from the integrated Poisson equation: \begin{equation}
    -2\pi G \Sigma (R,z) = K_z(R,z),
\end{equation} and an estimate for \rhodm can be obtained using Equation \ref{eq:rhodm_from_sigma}.

An alternative way of estimating the dark matter density in the midplane is by directly using Gauss' Law: given the fit vertical acceleration profile $K_z(z)$, the corresponding density in the midplane is then \begin{equation}
    \rho_0 = -\frac{1}{4\pi G}\dv{K_z}{z}\Big|_{z=0}.
\end{equation} By using estimates of the local volume density of the baryonic components, the density of dark matter in the midplane can then be calculated as \begin{equation} \label{eq:rho_dm_DAM}
    \rho_{0,\mathrm{DM}} = \rho_0 - \rho_{0,*} - \rho_{0,\mathrm{gas}},
\end{equation} which also requires existing knowledge of the stellar and gas volume densities in the midplane. This second method performs somewhat worse on average than using Equation \ref{eq:rhodm_from_sigma}, so we choose not to use it moving forward. 

Note that this procedure is not how \rhodm was estimated in previous work that used pulsar timing data to infer the value of \rhodm \citep{Chakrabarti2021,Donlon2024,Donlon2025}. We use a different approach for several reasons. First, \cite{Chakrabarti2021} and \cite{Donlon2024} used so-called ``$\alpha$ models'' to compute the local density of matter, which have since been shown to underestimate the local value of \rhodm \citep{Donlon2025}. Additionally, these papers each optimize a variety of potential models to the acceleration data directly, which is computationally expensive; this is fine when it only needs to be run once for those works, but it is infeasible to run many hundreds of times for the purposes of this study. Finally, it becomes difficult to perform unsupervised fits of potential models directly for small numbers of sources ($\lesssim20$), meaning that fits will often not provide meaningful constraints on the potential parameters. In contrast, the method provided here is computationally fast and performs well even for a small number of sources, making it advantageous in this particular use case. 

We explored adding additional terms to the direct acceleration calculation in order to explicitly capture disequilibrium effects (see Appendix \ref{app:diseq_on_dams} for a discussion of this). Ultimately this leads to a small improvement in the recovery of the local dark matter density, although it is somewhat more computationally intense. While we do not spend much time on this idea here, such considerations make sense for studies that wish to constrain properties of specific disequilibrium features, such as the interpretation of an asymmetry in the strength of $a_z$ above and below the disk plane by \cite{Donlon2025}.

\subsection{Recovery of Vertical Force}

\begin{figure*}
    \centering
    \includegraphics[width=\linewidth]{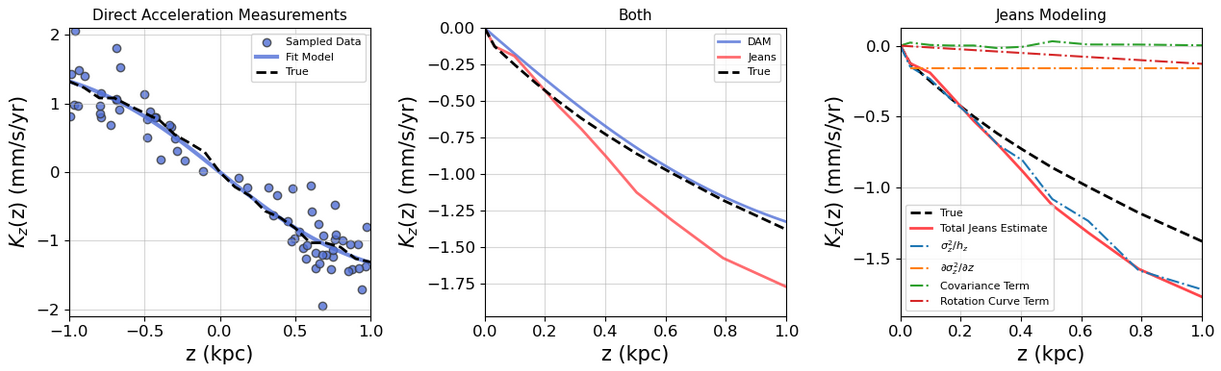}
    \caption{Recovery of the vertical force ($K_z$) using direct acceleration measurements (left), Jeans modeling (right), and a comparison of the two methods (center). Mock data for both methods are drawn from the interacting simulation across the region $8<r<9$ kpc and $15^\circ<\phi<30^\circ$; this volume corresponds to an underdensity between spiral arms, causing the Jeans modeling to fail. Although there is significant scatter in the individual direct acceleration measurements, our method is able to faithfully recover the true underlying vertical force curve. Jeans modeling overestimates the vertical force at heights beyond 300 pc from the midplane, leading to a significant error in the corresponding estimate of the local (dark matter) density in this region.}
    \label{fig:kz_comp}
\end{figure*}

Although both the Jeans modeling and direct acceleration frameworks we have established above will accurately recover the vertical force curve in an equilibrium environment, it is valuable to see how these two methods perform in a more realistic setting. In Figure \ref{fig:kz_comp}, we illustrate the ability of both methods to measure the vertical force curve of the interacting simulation. This test was run using mock data sampled from the region $8<r<9$ kpc and $15^\circ<\phi<30^\circ$, which corresponds to an especially underdense region between two major spiral arms in the interacting simulation. It is clear that, in this case, the DAM method significantly outperforms Jeans modeling at reproducing the actual underlying force curve of the simulated galaxy. 

This particular region was selected because it leads to a particularly poor performance of our Jeans modeling setup, which tends to fail in regions of low surface density. This illustrates the problems that can arise in Jeans modeling, and can be solved by using direct acceleration measurements. Note that in regions of the simulated disk which do not have low surface density, Jeans modeling performs roughly as well as the DAM method, yet only rarely outperforms DAM; in the next section, we explore in more detail why the performance of Jeans modeling is related to the local surface density.

\section{Recovery of Local Dark Matter Density}

\begin{figure}
    \centering
    \includegraphics[width=\linewidth]{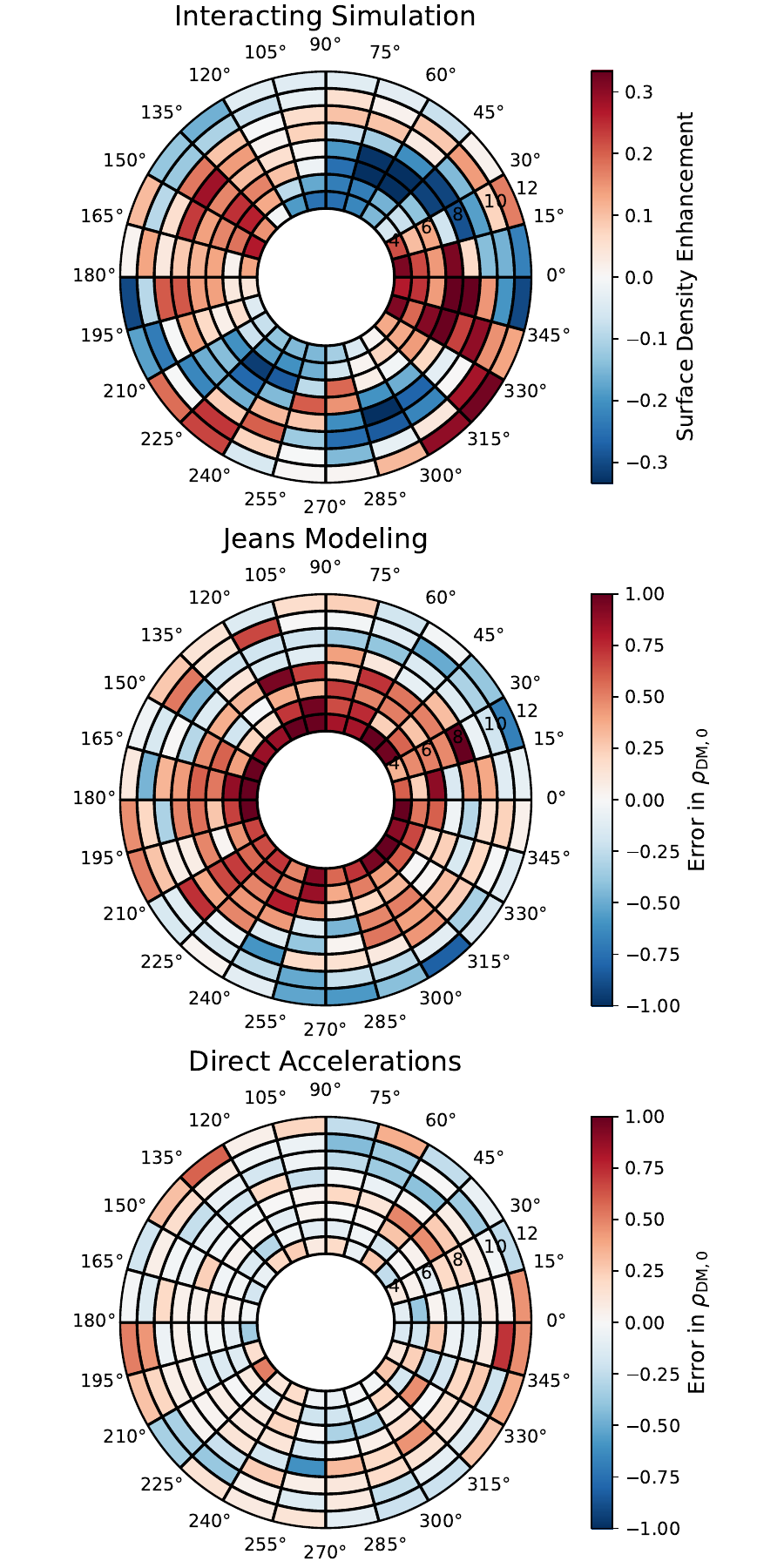}
    \caption{Face-on surface density and recovered dark matter density in the midplane (\rhodmnospace). \textit{Top:} Local surface density enhancement in the interacting simulation, painted onto a face-on projection of the host galaxy disk. Red (blue) corresponds to a higher (lower) surface density in that region compared to the average over a ring at the same Galactocentric radius; more saturated colors indicates a larger discrepancy. The spiral arms are clearly visible. \textit{Middle:} Fractional error in the recovered \rhodm when using Jeans modeling. Jeans modeling overestimates the value of \rhodm in many places, particularly in the locations in-between spiral arm overdensities. \textit{Bottom:} Fractional error in recovered values of \rhodm when using direct acceleration measurements. The error and bias in the density inferred from direct acceleration measurements is on average much smaller than those from Jeans modeling.}
    \label{fig:faceon_rhodm_err}
\end{figure}

\begin{figure*}
    \centering
    \includegraphics[width=\linewidth]{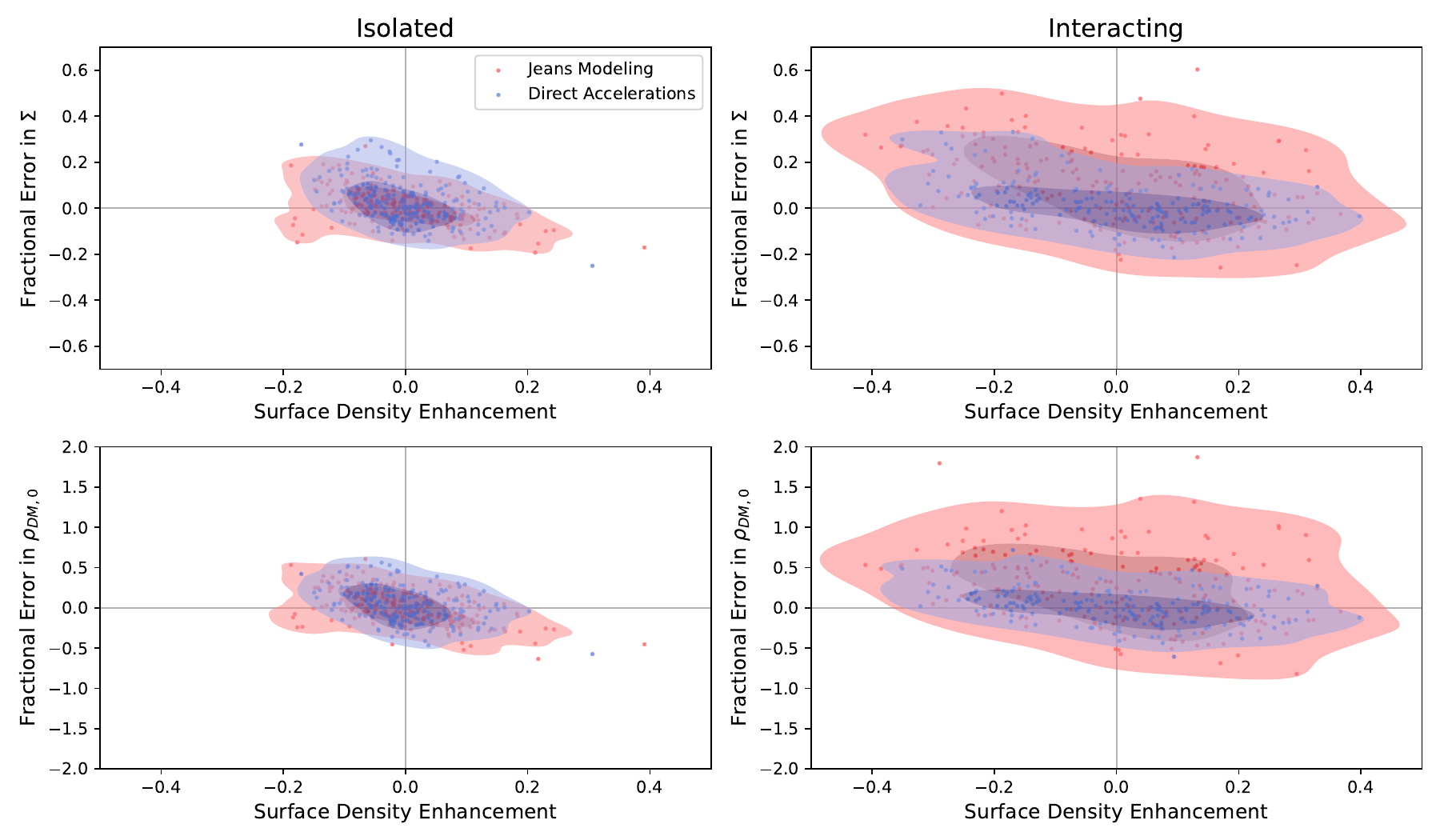}
    \caption{Fractional error in the recovered surface density and dark matter density in the midplane (\rhodmnospace) as a function of the local surface density enhancement ($\Delta\Sigma$). Results for the isolated simulation are shown in the left column, and results for the interacting simulation are provided in the right column. In general, the contours of the data trend from the upper left to the bottom right of each panel; this indicates that negative $\Delta\Sigma$ corresponds to an overestimate in both inferred surface density and \rhodmnospace, although this effect is larger for Jeans modeling (red) than it is for direct acceleration techniques (blue). The values of $\Delta\Sigma$ are small in the isolated simulation, and the corresponding errors are roughly equal for direct acceleration measurements and Jeans modeling. In the interacting simulation, the values of $\Delta\Sigma$ become much larger (due to more significant departures from dynamical equilibrium); while the errors in the inferred values are large for Jeans modeling, the errors are much smaller when using direct acceleration measurements.}
    \label{fig:scatter_rhodm_err}
\end{figure*}

\begin{figure*}
    \centering
    \includegraphics[width=\linewidth]{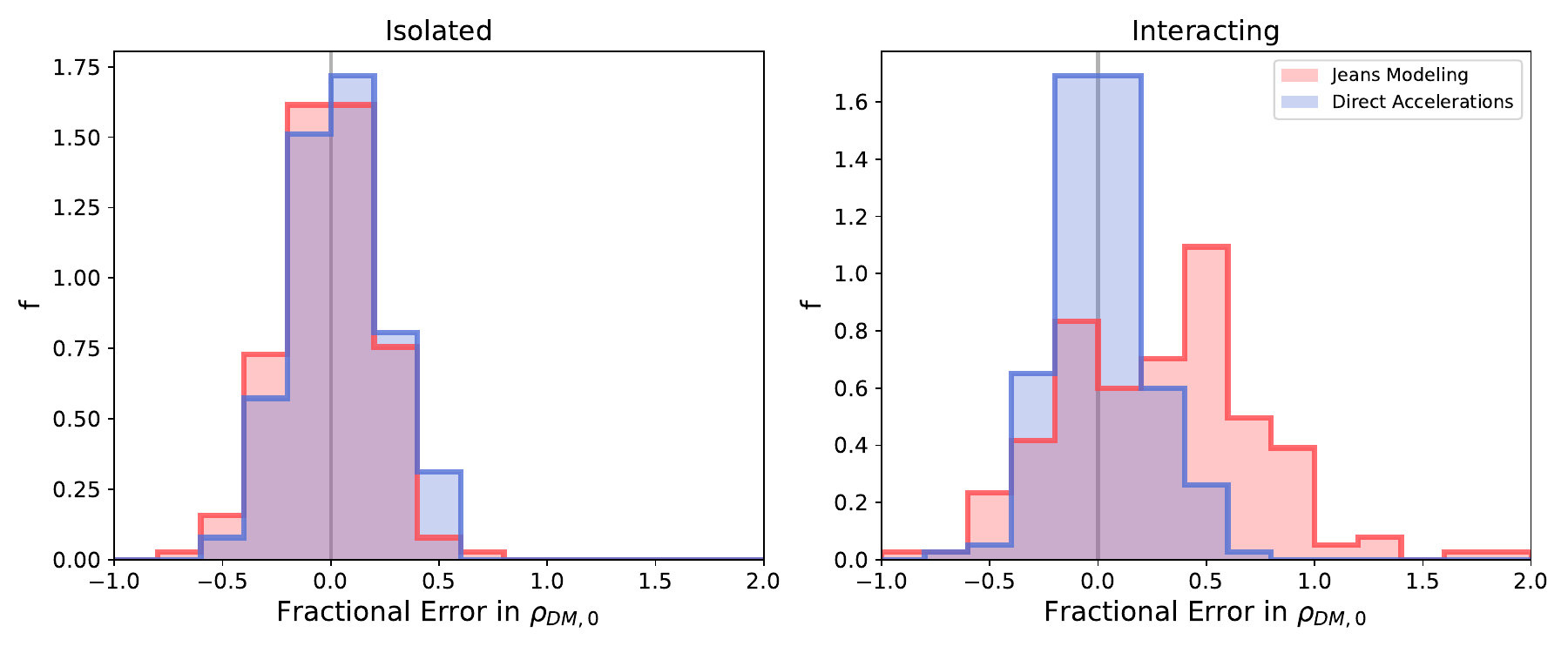}
    \caption{Error in \rhodm inferred from Jeans modeling and direct acceleration data in both simulations. In the isolated simulations, Jeans modeling and direct acceleration techniques perform equally as well. In the interacting simulation, Jeans modeling has a significant bias towards substantially overestimating the value of \rhodmnospace, while direct acceleration measurements perform at the same level as in the isolated simulation.}
    \label{fig:hist_rhodm_err}
\end{figure*}

\begin{figure*}
    \centering
    \includegraphics[width=\linewidth]{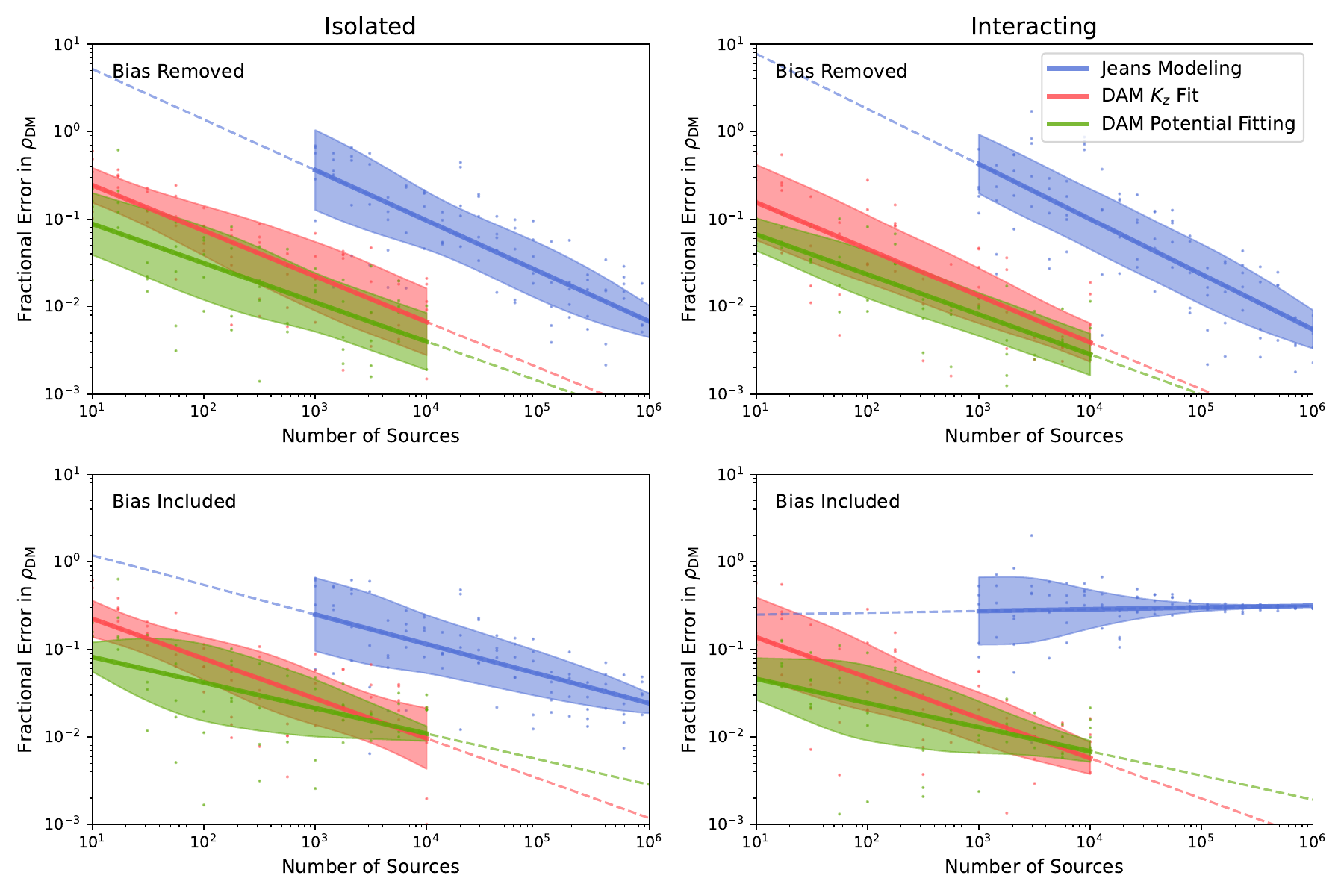}
    \caption{Fractional error in \rhodm for different measurement techniques as a function of the number of datapoints available (after removing biases in the inferred values). Each random pull from the simulation is shown as individual points, and a power-law fit to these points plus the 1-$\sigma$ scatter in the points are provided as a solid line and shaded region. The top row contains data where the biases in the inferred value of \rhodm have been removes, while the bottom row shows the results when the biases are left in the inferred values. For the same number of sources, direct acceleration potential modeling slightly outperforms direct acceleration $K_z$ profile modeling, which both dramatically outperform Jeans modeling. When the biases are removed, the direct acceleration techniques perform at roughly the same level as the Jeans modeling technique, but with 1000 times fewer sources. When bias is included, things are only slightly changed in the isolated simulation. However, in the interacting simulation, Jeans modeling has a significant bias for the recovered value of \rhodmnospace. As a result, in this case Jeans modeling never performs better than direct acceleration modeling regardless of the number of sources used. }
    \label{fig:unc_n}
\end{figure*}

Here we evaluate the performance of the kinematic and direct acceleration methods at recovering the local dark matter density in both the isolated and interacting simulations. In the isolated simulation, the behavior of the disk is essentially time-independent, and the assumption of axisymmetry made above is approximately satisfied. However, in the case of the interacting simulation, the large amount of disequilibrium in the disk violates the assumptions of time-independence and axisymmetry, which may lead to incorrect estimates of \rhodmnospace. 

\subsection{Spatial Variation and Surface Density Bias} \label{sec:surf_dens_bias}

\cite{HainesDonghia2019} pointed out that Jeans modeling tends to overestimate the surface density of the disk in regions of low mean density. In order to test this claim in our simulations, we define the ``local surface density enhancement'' to be \begin{equation}
    \Delta\Sigma(R,z,\phi) = \frac{\Sigma(R,z,\phi) - \bar{\Sigma}_\mathrm{ring}(R,z)}{\bar{\Sigma}_\mathrm{ring}(R,z)},
\end{equation} where the mean surface density over a ring with Galactocentric radius $R$ and height $|z|$ from the midplane is \begin{align}
    \bar{\Sigma}_\mathrm{ring}(R,z) =& \\ \nonumber
    \frac{1}{2\pi \Delta R}& \int_0^{2\pi} \int_0^z \int_R^{R + \Delta R} \rho(R', z', \phi) \; \dd R' \; \dd z' \; \dd\phi.
\end{align} $\Delta\Sigma(R,z,\phi)$ is equal to the fractional difference between the surface density of a small section of the disk and the mean surface density across the entire disk at that radius. 

We evaluate the value of \rhodm using both the kinematic and direct acceleration method for many points on the disk. Each evaluation consists of stars within a region spanning 7.5$^\circ$ in $\phi$, 1 kpc in $R$, and 2 kpc in $z$ (1 kpc on either side of the midplane). We only evaluate \rhodm between 4 kpc $<$ $R$ $<$ 12 kpc, because inside 4 kpc the bar/bulge dominates the dynamics of the galaxy, and outside 12 kpc there are too few stars to do meaningful Jeans analysis\footnote{This is a potentially powerful advantage of direct acceleration measurements; if accelerations can be measured for a couple dozen sources at large radii -- which may become possible in the future -- we could in principle constrain the dynamics of the disk at radii where Jeans modeling is not practical due to a lack of data.}. The number of stars in each region varied in each region because the surface density of the disk decreases as $R$ increases, but the mean number of stars in a region was roughly 6000 stars. For the direct acceleration measurements, we always used 50 sources \citep[comparable to the total number of available pulsar accelerations today,][]{Donlon2025}. These sources were randomly sampled from a uniform distribution within a 1 kpc box centered on the region of interest.  

The inferred value of \rhodm is then compared to the true dark matter density in the simulation. Both methods recover the value of \rhodm with the same level of accuracy in the isolated simulation. There is a large discrepancy between the two methods in the interacting simulation, however. This is shown across the surface of the disk in Figure \ref{fig:faceon_rhodm_err}; the direct acceleration method recovers \rhodm to a high level of precision even in the interacting disk, while Jeans modeling shows larger errors and bias in its recovery of the local dark matter density. Jeans modeling particularly struggles in the inner galaxy, and in regions of lower local surface density, which is consistent with the findings of \cite{HainesDonghia2019}. Quantitatively, across underdense regions with $\Delta\Sigma < -0.2$ in the interacting simulation, Jeans modeling overestimates \rhodm with a median bias of 51\% and a standard deviation of 36\%, while the DAM method recovers \rhodm with a median bias of just 18\% and a standard deviation of 19\%.

We show these results as a function of the local surface density enhancement in Figure \ref{fig:scatter_rhodm_err}. In the isolated simulation, Jeans modeling and direct acceleration measurements recover both the surface density and the local dark matter density equally well. In the interacting simulation, however, there is a clear difference in the performance of the two methods. Direct acceleration modeling recovers both the surface density and the value of \rhodm equally well in both the isolated and interacting case, indicating its resilience to disequilibrium effects. Conversely, Jeans modeling performs markedly worse in the interacting simulation than in the isolated simulation, indicating that this method becomes unreliable in a realistic scenario where the stellar disk is not in equilibrium. In general, Jeans modeling overestimates the surface density and local dark matter density across the entire disk, but this issue is most pronounced in regions of locally low surface density. 

There is a small feature in the top-right panel of Figure \ref{fig:scatter_rhodm_err} where the DAM method also overestimates the local surface density for a small number of regions in the disk with negative surface density enhancement. Analysis indicates that these large errors arise in parts of the disk just beyond the cusp of a spiral arm, where the local acceleration field is not well approximated by an exponential profile. However, there are only a few regions across the disk where this is a problem, and while it leads to a substantial error in recovering the surface density, it only marginally increases the error in recovering the local dark matter density. Additionally, visual inspection of the acceleration data in these regions makes it clear that use of an exponential profile is not appropriate, making it unlikely that this type of problem would arise in actual data. 

The performance of both methods at recovering \rhodm is also shown as a histogram in Figure \ref{fig:hist_rhodm_err}. Jeans modeling and the direct acceleration method perform equally well in the isolated simulation, with standard deviations of 0.22 in fractional error for both methods. In the interacting case, the direct acceleration method performed just as well with the same scatter of 0.22, but Jeans modeling performed significantly worse with a standard deviation of 0.44 in fractional error. 

In this analysis, we assume that we know the Sun's distance from the midplane exactly, and that there are no local perturbations to the acceleration field (in other words, a peculiar Solar acceleration that is not felt by observed pulsars). We test how the direct acceleration procedure performs at determining the midplane DM density when these assumptions are not true in Appendix \ref{app:z0_aphi_err}. We find that there is essentially no difference in the recovery of $\rho_{DM,0}$ when one includes reasonable observational uncertainties on the vertical position of the Sun and a peculiar Solar acceleration. Direct accelerations are remarkably resilient in this case -- even significant errors in these values only lead to small reductions in the method's ability to recover $\rho_{DM,0}$. 

\subsection{Precision per Source} \label{sec:precision_per_source}

We have established that Jeans modeling performs worse than direct acceleration methods at recovering the correct value of \rhodmnospace. As per the central question of this study -- ``what is the value of a single acceleration measurement'' -- it is also useful to determine the relative performance of Jeans modeling and direct acceleration measurements as a function of the number of sources available. 

In order to test this, we randomly sample the disk many times for both methods. Each test consisted of randomly selecting a point on the disk within 4 kpc $<$ $R$ $<$ 12 kpc and 0 $<$ $\phi$ $<$ 2$\pi$, and then generating the sample of data. In the case of direct accelerations, we sample the required number of sources from a 1 kpc box centered on this point (similar to the previous procedure). Since Jeans modeling requires a large number of stars, for the kinematic data we include all stars within 1 kpc in $R$ centered on the point of interest, and a range of 7.5$^\circ$ in $\phi$. If this selection included more stars than necessary for a given test, we randomly selected stars from this sample until we had the required number of datapoints. If this selection did not include enough stars (this was the case for the tests that included as many as 1 million stars), we expanded the range of the selection in $\phi$ until enough star particles were present in the selection. This was done 10 times for each number of sources being tested. For the direct acceleration measurements, we tested between $10^1$ and $10^4$ sources, while for the kinematic data we tested between $10^3$ and $10^6$ sources. We tested different ranges of the number of sources for the different methods because of the availability of the observed data in each case, combined with the complexity of each method as the number of sources increased. 

Here, we also include the results of fitting a simple two-component potential model to the direct acceleration data, consisting of an exponential disk and an NFW halo potential \citep{NFW}. This procedure is similar to the procedures of the studies which estimate \rhodm from pulsar acceleration observations \citep{Chakrabarti2021,Donlon2024,Donlon2025}. The potential fitting method performs somewhat better than the procedure from Section \ref{sec:direct_acc_modeling}, although it has the problem of not always constraining the parameters of the potential for $N<100$. In other words, potential fitting outperforms fitting a profile to the de-projected vertical accelerations, but only when it produces a valid constraint, which is not always the case. 

The results of these tests are shown in Figure \ref{fig:unc_n}. The most striking result is for the interacting simulation with biases included: Jeans modeling never converges to the correct value of \rhodm regardless of how many stars are used, because the underlying kinematic distribution retains the imprint of the satellite perturbation. Direct acceleration methods, by contrast, continue to improve as sources are added and are unaffected by the disequilibrium. Note that each method has a systematic bias in the recovery of \rhodmnospace (in other words, the inferred dark matter density asymptotically approaches a value as one adds more sources); we show the results with these biases included and removed. 

When the biases are not considered, Jeans modeling and direct acceleration measurements are offset by a factor of $\sim$100 for the $K_z$ fitting method, and a factor of $\sim$1000 for the potential fitting method. This affirms our theoretical baseline that each acceleration measurement contains equivalent information to thousands of stars. We emphasize that this is the result in a setting without any disequilibrium processes, selection effects, measurement uncertainties, or other systematic issues that could lead to problems recovering the local dark matter density; at the most basic level, direct acceleration measurements outperform kinematic measurements by a few orders of magnitude, and this result is not caused by improper measurements or by ignoring confounding variables. 

When the biases are included for the interacting simulation, however, Jeans modeling is unable to correctly infer the local dark matter density to high precision no matter how many sources are used. This is because Jeans modeling severely overestimates the value of \rhodmnospace in disequilibrium settings. On the other hand, the precision of the direct acceleration methods continues to improve as additional sources are added at least until 10$^4$ sources. This once again highlights the advantage of using direct acceleration methods rather than kinematic methods in disks which are significantly out of equilibrium, such as the MW disk. Crucially, this point is why the precision of the inferred local dark matter density does not appear to scale with the number of sources in Figure \ref{fig:unc_n}; the bias in kinematic methods begins to dominate the possible precision of these studies.

\section{Discussion}

\begin{figure}
    \centering
    \includegraphics[width=\linewidth]{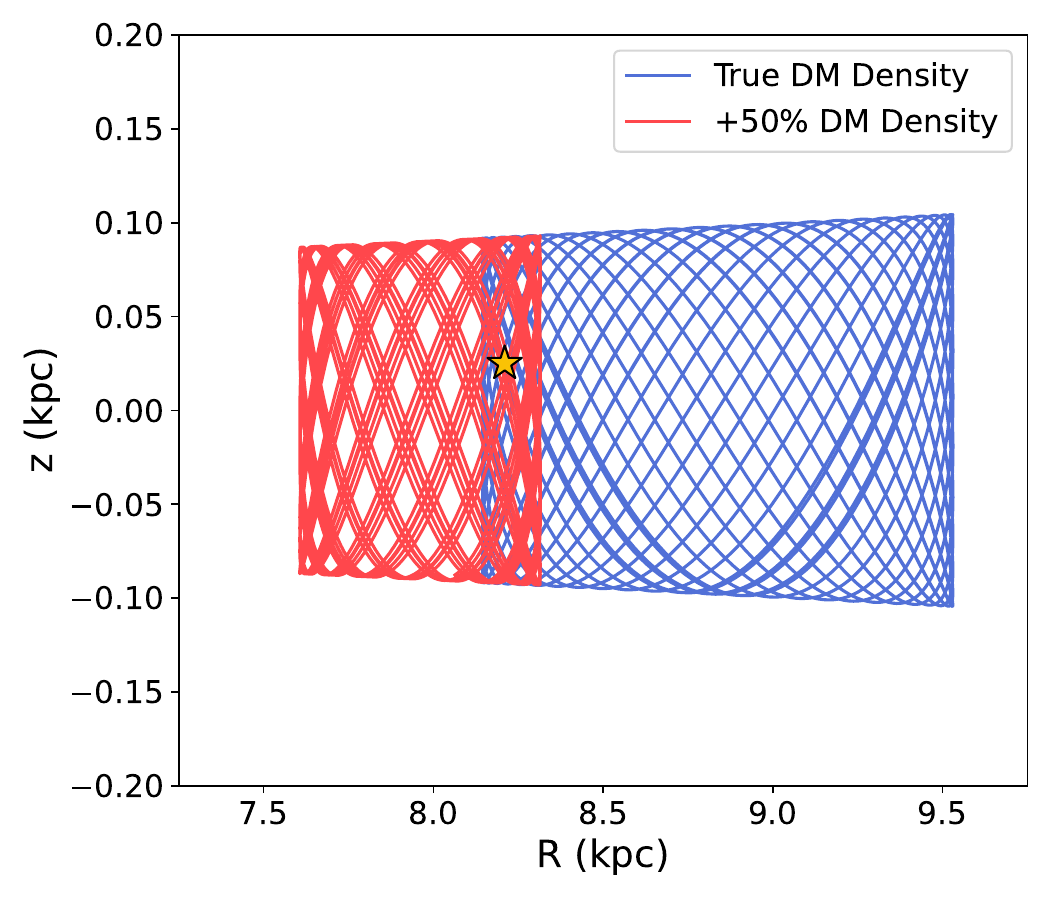}
    \caption{The orbit of the Sun in a static potential (current location marked by a gold star). The blue orbit shows the Solar orbit for the value of $\rho_{DM,0}$ used in the Gala MilkyWayPotential2022 model (which is similar to the value obtained from the pulsar direct acceleration measurements). The red orbit shows the Solar orbit if one assumes a value of $\rho_{DM,0}$ that is 50\% too large (a realistic amount of bias for Jeans modeling based on our analysis). The stark difference between the two orbits highlights how assuming an incorrect local density of dark matter can significantly change the inferred dynamics of disk stars.}
    \label{fig:sun_orbit}
\end{figure}

This study establishes two complementary conclusions. Even in an equilibrium disk, direct acceleration measurements carry an intrinsic per-source information advantage of orders of magnitude over kinematic tracers, because each acceleration constrains the local gravitational field directly rather than through the time-averaged distribution function. In a disequilibrium disk, this advantage compounds: Jeans-based methods develop a systematic bias -- a median overestimate of \rhodm of $\sim$50\% in underdense regions of the interacting simulation -- that cannot be reduced by increasing the stellar sample, while direct acceleration methods remain unaffected.

If the local dark matter densities from Jeans modeling are taken to be true, we are likely overestimating the value of \rhodm by a substantial amount. Figure~\ref{fig:sun_orbit} shows the effect of this on the inferred orbit of the Sun. If the presumed value of the local dark matter density is 50\% too large (which is consistent with our results given the theoretical performance of Jeans modeling in this study), the inferred Solar orbit would be dramatically different from its true orbit throughout the MW. This has profound implications for studies of the MW, many of which compute orbits and/or orbital parameters for many stars in a model potential that is fit to kinematic observations, and rely on these quantities for the key results of their work. Additionally, studies which attempt to directly detect dark matter \citep{Lewin1996,Feng2010,Freese2013} rely on a presumed dark matter flux, which itself depends on the local dark matter density; the standard halo model used by many of these studies has been built from decades of Jeans modeling results, and may therefore result in predictions for dark matter scattering signals which are too large. The bias present in surface density and local dark matter density inferred through Jeans methods and other similar techniques warrants careful consideration.

\section{Conclusion}

It has recently become possible to measure fundamental properties of the Milky Way by directly measuring the accelerations of objects near the Sun, rather than relying on the positions and velocities of stars. This includes measurements of the local dark matter density. The most recent direct acceleration study produced a measurement of \rhodm that is competitive with existing kinematic constraints, warranting investigation into this novel method. However, the overall performance of direct acceleration studies compared to classical kinematic methods is currently unknown. 

We analyze various studies in the literature which measure the local dark matter density, and show that there is a modest but systematic tension in the inferred value of \rhodm between two clusters (types) of studies. The first cluster includes studies based on the rotation curve, synthesis/review studies, and pulsar acceleration studies, which produce a relatively low value for \rhodm compared to studies that rely on Jeans modeling, distribution function fitting, the phase space spiral, or tidal streams. This suggests the existence of a bias in one or both of these groups of studies, leading to an incorrect estimate for the local dark matter density. 

We assess the performance of Jeans modeling and direct acceleration methods in two simulated galaxies: an isolated galaxy that is roughly in dynamical equilibrium, and a galaxy that is interacting with a Sgr-like dwarf galaxy satellite which is significantly out of equilibrium. There is a substantial body of evidence that the MW disk is more similar to the interacting case in that assumptions of time-independence and axisymmetry in the distribution function do not hold. We show that while Jeans modeling performs well in the isolated (equilibrium) case, it severely overestimates the surface density and local dark matter density in the interacting simulation. On the other hand, direct acceleration methods do not have this issue, and continue to effectively constrain the local dark matter density even in the presence of substantial disequilibrium. 

We also assess the relative performance of kinematic methods and direct acceleration methods, both in the literature and in the simulations, in order to determine how many acceleration measurements are required to match the precision and accuracy of kinematic methods. Based on the literature, a single acceleration measurement contains the same amount of information as $10^3-10^7$ stars, depending on the specific study. We also show that kinematic measurements appear to be asymptotically approaching a limit in their ability to estimate the value of \rhodmnospace, and direct acceleration studies may begin to outperform kinematic/dynamical studies with as few as 200 observed accelerations. Based on the theoretical performance of Jeans modeling and direct acceleration methods in the simulated data, direct accelerations are able to accurately recover the true value of \rhodm even when the disk is significantly out of equilibrium, whereas Jeans modeling has a significant bias towards a too-high value of \rhodmnospace. 

We emphasize that an incorrect inference of the local dark matter density can have significant implications for a wide range of kinematic and dynamical studies, which often rely on the calculation of orbital parameters from a presumed potential model. If the value of the local dark matter density is taken from Jeans modeling results, it is likely that we are overestimating the contribution of dark matter to accelerations near the disk. This could result in incorrect inferred parameters including orbital actions, guiding radii, and others, which could have serious impacts on the results of studies which rely on such quantities. 

Overall, we show that direct acceleration measurements are particularly valuable compared to kinematic data of stars, both in the relative information available in each case, as well as the ability of direct acceleration methods to correctly recover the underlying value of \rhodm. In the future, when planning new missions and the priority of obtaining different kinds of data, we emphasize the importance of considering the value of direct acceleration measurements and their close synergy with exoplanet and time-domain techniques.

\acknowledgments

We would like to thank Alice Quillen and Michael Lam for helpful discussions. 
TD acknowledges partial support from the NSF award OAC 2513443. SC acknowledges support from NASA EPSCoR CAN AL-80NSSC24M0104, STSCI GO 17505, and the Margaret Burbidge fellowship. ED acknowledges the Vilas Midcareer Award.

\bibliographystyle{aasjournal}
\bibliography{references.bib}

@ARTICLE{DamourTaylor1992,
       author = {{Damour}, Thibault and {Taylor}, J.~H.},
        title = "{Strong-field tests of relativistic gravity and binary pulsars}",
      journal = {\prd},
         year = 1992,
        month = mar,
       volume = {45},
       number = {6},
        pages = {1840-1868},
          doi = {10.1103/PhysRevD.45.1840},
       adsurl = {https://ui.adsabs.harvard.edu/abs/1992PhRvD..45.1840D}
}

@ARTICLE{Arora2026,
       author = {{Arora}, Arpit and {Ferguson}, Peter S. and {Nibauer}, Jacob and {Shipp}, Nora and {Reddy}, Videep and {Vasiliev}, Eugene and {Kohm}, Jack and {Marin}, Laurella C. and {Price-Whelan}, Adrian M. and {Erkal}, Denis and {Pearson}, Sarah and {Wetzel}, Andrew and {Bailin}, Jeremy and {Feldmann}, Robert},
        title = "{No Stream Left Unscathed: The imprint of a host galaxy}",
      journal = {arXiv e-prints},
         year = 2026,
        month = may,
          eid = {arXiv:2605.16200},
        pages = {arXiv:2605.16200},
          doi = {10.48550/arXiv.2605.16200},
archivePrefix = {arXiv},
       eprint = {2605.16200},
 primaryClass = {astro-ph.GA},
       adsurl = {https://ui.adsabs.harvard.edu/abs/2026arXiv260516200A}
}

@ARTICLE{Chakrabarti2022whitepaper,
       author = {{Chakrabarti}, Sukanya and {Drlica-Wagner}, Alex and {Li}, Ting S. and {Sehgal}, Neelima and {Simon}, Joshua D. and {Birrer}, Simon and {Brown}, Duncan A. and {Bernstein}, Rebecca and {Bolatto}, Alberto D. and {Chang}, Philip and {Dawson}, Kyle and {Demorest}, Paul and {Grin}, Daniel and {Kaplan}, David L. and {Lazio}, Joseph and {Marshall}, Jennifer and {Murphy}, Eric J. and {Ransom}, Scott and {Robertson}, Brant E. and {Singh}, Rajeev and {Slosar}, An{\v{z}}e and {Treu}, Tommaso and {Tsai}, Yu-Dai and {Williams}, Benjamin F.},
        title = "{Snowmass2021 Cosmic Frontier White Paper: Observational Facilities to Study Dark Matter}",
      journal = {arXiv e-prints},
         year = 2022,
        month = mar,
          eid = {arXiv:2203.06200},
        pages = {arXiv:2203.06200},
          doi = {10.48550/arXiv.2203.06200},
archivePrefix = {arXiv},
       eprint = {2203.06200},
 primaryClass = {astro-ph.CO},
       adsurl = {https://ui.adsabs.harvard.edu/abs/2022arXiv220306200C}
}

@ARTICLE{Drlica-Wagner2022,
       author = {{Drlica-Wagner}, Alex and {Prescod-Weinstein}, Chanda and {Yu}, Hai-Bo and {Albert}, Andrea and {Amin}, Mustafa and {Banerjee}, Arka and {Baryakhtar}, Masha and {Bechtol}, Keith and {Bird}, Simeon and {Birrer}, Simon and {Bringmann}, Torsten and {Caputo}, Regina and {Chakrabarti}, Sukanya and {Chen}, Thomas Y. and {Croon}, Djuna and {Cyr-Racine}, Francis-Yan and {Dawson}, William A. and {Dvorkin}, Cora and {Gluscevic}, Vera and {Gilman}, Daniel and {Grin}, Daniel and {Hlo{\v{z}}ek}, Ren{\'e}e and {Leane}, Rebecca K. and {Li}, Ting S. and {Mao}, Yao-Yuan and {Meyers}, Joel and {Mishra-Sharma}, Siddharth and {Mu{\~n}oz}, Julian B. and {Munshi}, Ferah and {Nadler}, Ethan O. and {Parikh}, Aditya and {Perez}, Kerstin and {Peter}, Annika H.~G. and {Profumo}, Stefano and {Schutz}, Katelin and {Sehgal}, Neelima and {Simon}, Joshua D. and {Sinha}, Kuver and {Valluri}, Monica and {Wechsler}, Risa H.},
        title = "{Report of the Topical Group on Cosmic Probes of Dark Matter for Snowmass 2021}",
      journal = {arXiv e-prints},
         year = 2022,
        month = sep,
          eid = {arXiv:2209.08215},
        pages = {arXiv:2209.08215},
          doi = {10.48550/arXiv.2209.08215},
archivePrefix = {arXiv},
       eprint = {2209.08215},
 primaryClass = {hep-ph},
       adsurl = {https://ui.adsabs.harvard.edu/abs/2022arXiv220908215D}
}

@ARTICLE{Poggio2025,
       author = {{Poggio}, E. and {Khanna}, S. and {Drimmel}, R. and {Zari}, E. and {D'Onghia}, E. and {Lattanzi}, M.~G. and {Palicio}, P.~A. and {Recio-Blanco}, A. and {Thulasidharan}, L.},
        title = "{The great wave: Evidence of a large-scale vertical corrugation propagating outwards in the Galactic disc}",
      journal = {\aap},
         year = 2025,
        month = jul,
       volume = {699},
          eid = {A199},
        pages = {A199},
          doi = {10.1051/0004-6361/202451668},
archivePrefix = {arXiv},
       eprint = {2407.18659},
 primaryClass = {astro-ph.GA},
       adsurl = {https://ui.adsabs.harvard.edu/abs/2025A&A...699A.199P}
}

@ARTICLE{Lewin1996,
       author = {{Lewin}, J.~D. and {Smith}, P.~F.},
        title = "{Review of mathematics, numerical factors, and corrections for dark matter experiments based on elastic nuclear recoil}",
      journal = {Astroparticle Physics},
         year = 1996,
        month = dec,
       volume = {6},
       number = {1},
        pages = {87-112},
          doi = {10.1016/S0927-6505(96)00047-3},
       adsurl = {https://ui.adsabs.harvard.edu/abs/1996APh.....6...87L}
}

@article{Freese2013,
  title = {Colloquium: Annual modulation of dark matter},
  author = {Freese, Katherine and Lisanti, Mariangela and Savage, Christopher},
  journal = {Rev. Mod. Phys.},
  volume = {85},
  issue = {4},
  pages = {1561--1581},
  numpages = {0},
  year = {2013},
  month = {Nov},
  publisher = {American Physical Society},
  doi = {10.1103/RevModPhys.85.1561},
  url = {https://link.aps.org/doi/10.1103/RevModPhys.85.1561}
}

@ARTICLE{Chakrabarti_Blitz2011,
       author = {{Chakrabarti}, Sukanya and {Blitz}, Leo},
        title = "{Tidal Imprints of a Dark Sub-halo on the Outskirts of the Milky Way. II. Perturber Azimuth}",
      journal = {\apj},
         year = 2011,
        month = apr,
       volume = {731},
       number = {1},
          eid = {40},
        pages = {40},
          doi = {10.1088/0004-637X/731/1/40},
archivePrefix = {arXiv},
       eprint = {1007.1982},
 primaryClass = {astro-ph.GA},
       adsurl = {https://ui.adsabs.harvard.edu/abs/2011ApJ...731...40C}
}

@BOOK{LorimerKramer2012Handbook,
       author = {{Lorimer}, D.~R. and {Kramer}, M.},
        title = "{Handbook of Pulsar Astronomy}",
         year = 2012,
       adsurl = {https://ui.adsabs.harvard.edu/abs/2012hpa..book.....L},
      publisher = {Cambridge University Press}
}

@ARTICLE{Bonacaetal2020,
       author = {{Bonaca}, Ana and {Conroy}, Charlie and {Hogg}, David W. and {Cargile}, Phillip A. and {Caldwell}, Nelson and {Naidu}, Rohan P. and {Price-Whelan}, Adrian M. and {Speagle}, Joshua S. and {Johnson}, Benjamin D.},
        title = "{High-resolution Spectroscopy of the GD-1 Stellar Stream Localizes the Perturber near the Orbital Plane of Sagittarius}",
      journal = {\apjl},
         year = 2020,
        month = apr,
       volume = {892},
       number = {2},
          eid = {L37},
        pages = {L37},
          doi = {10.3847/2041-8213/ab800c},
archivePrefix = {arXiv},
       eprint = {2001.07215},
 primaryClass = {astro-ph.GA},
       adsurl = {https://ui.adsabs.harvard.edu/abs/2020ApJ...892L..37B}
}

@ARTICLE{Chakrabartietal2026,
       author = {{Chakrabarti}, Sukanya and {Chang}, Philip and {Profumo}, Stefano and {Craig}, Peter},
        title = "{Constraints on a Dark Matter Subhalo Near the Sun from Pulsar Timing}",
      journal = {\prl},
         year = 2026,
        month = jan,
       volume = {136},
       number = {4},
          eid = {041201},
        pages = {041201},
          doi = {10.1103/29xz-nt5z},
archivePrefix = {arXiv},
       eprint = {2507.16932},
 primaryClass = {astro-ph.GA},
       adsurl = {https://ui.adsabs.harvard.edu/abs/2026PhRvL.136d1201C}
}

@ARTICLE{Donghia16,
       author = {{D'Onghia}, E. and {Madau}, P. and {Vera-Ciro}, C. and {Quillen}, A. and {Hernquist}, L.},
        title = "{Excitation of Coupled Stellar Motions in the Galactic Disk by Orbiting Satellites}",
      journal = {\apj},
         year = 2016,
        month = may,
       volume = {823},
       number = {1},
          eid = {4},
        pages = {4},
          doi = {10.3847/0004-637X/823/1/4},
archivePrefix = {arXiv},
       eprint = {1511.01503},
 primaryClass = {astro-ph.GA},
       adsurl = {https://ui.adsabs.harvard.edu/abs/2016ApJ...823....4D}
}

@ARTICLE{Donghia2010,
       author = {{D'Onghia}, Elena and {Vogelsberger}, Mark and {Faucher-Giguere}, Claude-Andre and {Hernquist}, Lars},
        title = "{Quasi-resonant Theory of Tidal Interactions}",
      journal = {\apj},
         year = 2010,
        month = dec,
       volume = {725},
       number = {1},
        pages = {353-368},
          doi = {10.1088/0004-637X/725/1/353},
archivePrefix = {arXiv},
       eprint = {1009.3927},
 primaryClass = {astro-ph.CO},
       adsurl = {https://ui.adsabs.harvard.edu/abs/2010ApJ...725..353D}
}

@ARTICLE{Bovy2017b,
       author = {{Bovy}, Jo},
        title = "{Stellar inventory of the solar neighbourhood using Gaia DR1}",
      journal = {\mnras},
         year = 2017,
        month = sep,
       volume = {470},
       number = {2},
        pages = {1360-1387},
          doi = {10.1093/mnras/stx1277},
archivePrefix = {arXiv},
       eprint = {1704.05063},
 primaryClass = {astro-ph.GA},
       adsurl = {https://ui.adsabs.harvard.edu/abs/2017MNRAS.470.1360B}
}

@ARTICLE{Putney2025,
       author = {{Putney}, Eric and {Shih}, David and {Lim}, Sung Hak and {Buckley}, Matthew R.},
        title = "{ClearPotential: Revealing Local Dark Matter in Three Dimensions}",
      journal = {arXiv e-prints},
         year = 2025,
        month = dec,
          eid = {arXiv:2512.09989},
        pages = {arXiv:2512.09989},
          doi = {10.48550/arXiv.2512.09989},
archivePrefix = {arXiv},
       eprint = {2512.09989},
 primaryClass = {astro-ph.GA},
       adsurl = {https://ui.adsabs.harvard.edu/abs/2025arXiv251209989P}
}

@ARTICLE{Donlon2026,
       author = {{Donlon}, II, Thomas and {Widrow}, Lawrence M. and {Chakrabarti}, Sukanya},
        title = "{Mean mass density near the Sun from the divergence theorem and pulsar accelerations}",
      journal = {\prd},
         year = 2026,
        month = mar,
       volume = {113},
       number = {6},
          eid = {063033},
        pages = {063033},
          doi = {10.1103/3g5b-cq14},
archivePrefix = {arXiv},
       eprint = {2511.15865},
 primaryClass = {astro-ph.GA},
       adsurl = {https://ui.adsabs.harvard.edu/abs/2026PhRvD.113f3033D}
}

@ARTICLE{Craig2023,
       author = {{Craig}, Peter and {Chakrabarti}, Sukanya and {Sanderson}, Robyn E. and {Nikakhtar}, Farnik},
        title = "{Building an Acceleration Ladder with Tidal Streams and Pulsar Timing}",
      journal = {\apjl},
         year = 2023,
        month = mar,
       volume = {945},
       number = {2},
          eid = {L32},
        pages = {L32},
          doi = {10.3847/2041-8213/acba15},
archivePrefix = {arXiv},
       eprint = {2211.00613},
 primaryClass = {astro-ph.GA},
       adsurl = {https://ui.adsabs.harvard.edu/abs/2023ApJ...945L..32C}
}

@ARTICLE{Price-Whelan2021,
       author = {{Price-Whelan}, Adrian M. and {Hogg}, David W. and {Johnston}, Kathryn V. and {Ness}, Melissa K. and {Rix}, Hans-Walter and {Beaton}, Rachael L. and {Brownstein}, Joel R. and {Garc{\'\i}a-Hern{\'a}ndez}, D.~A. and {Hasselquist}, Sten and {Hayes}, Christian R. and {Lane}, Richard R. and {Shetrone}, Matthew and {Sobeck}, Jennifer and {Zasowski}, Gail},
        title = "{Orbital Torus Imaging: Using Element Abundances to Map Orbits and Mass in the Milky Way}",
      journal = {\apj},
         year = 2021,
        month = mar,
       volume = {910},
       number = {1},
          eid = {17},
        pages = {17},
          doi = {10.3847/1538-4357/abe1b7},
archivePrefix = {arXiv},
       eprint = {2012.00015},
 primaryClass = {astro-ph.GA},
       adsurl = {https://ui.adsabs.harvard.edu/abs/2021ApJ...910...17P}
}

@ARTICLE{Gaskins2016,
       author = {{Gaskins}, Jennifer M.},
        title = "{A review of indirect searches for particle dark matter}",
      journal = {Contemporary Physics},
         year = 2016,
        month = oct,
       volume = {57},
       number = {4},
        pages = {496-525},
          doi = {10.1080/00107514.2016.1175160},
archivePrefix = {arXiv},
       eprint = {1604.00014},
 primaryClass = {astro-ph.HE},
       adsurl = {https://ui.adsabs.harvard.edu/abs/2016ConPh..57..496G}
}

@ARTICLE{TulinYu2018,
       author = {{Tulin}, Sean and {Yu}, Hai-Bo},
        title = "{Dark matter self-interactions and small scale structure}",
      journal = {\physrep},
         year = 2018,
        month = feb,
       volume = {730},
        pages = {1-57},
          doi = {10.1016/j.physrep.2017.11.004},
archivePrefix = {arXiv},
       eprint = {1705.02358},
 primaryClass = {hep-ph},
       adsurl = {https://ui.adsabs.harvard.edu/abs/2018PhR...730....1T}
}

@ARTICLE{Feng2010,
       author = {{Feng}, Jonathan L.},
        title = "{Dark Matter Candidates from Particle Physics and Methods of Detection}",
      journal = {\araa},
         year = 2010,
        month = sep,
       volume = {48},
        pages = {495-545},
          doi = {10.1146/annurev-astro-082708-101659},
archivePrefix = {arXiv},
       eprint = {1003.0904},
 primaryClass = {astro-ph.CO},
       adsurl = {https://ui.adsabs.harvard.edu/abs/2010ARA&A..48..495F}
}

@ARTICLE{SpringelHernquist2003,
       author = {{Springel}, Volker and {Hernquist}, Lars},
        title = "{Cosmological smoothed particle hydrodynamics simulations: a hybrid multiphase model for star formation}",
      journal = {\mnras},
         year = 2003,
        month = feb,
       volume = {339},
       number = {2},
        pages = {289-311},
          doi = {10.1046/j.1365-8711.2003.06206.x},
archivePrefix = {arXiv},
       eprint = {astro-ph/0206393},
 primaryClass = {astro-ph},
       adsurl = {https://ui.adsabs.harvard.edu/abs/2003MNRAS.339..289S}
}

@ARTICLE{Hopkins2013,
       author = {{Hopkins}, Philip F.},
        title = "{A general class of Lagrangian smoothed particle hydrodynamics methods and implications for fluid mixing problems}",
      journal = {\mnras},
         year = 2013,
        month = feb,
       volume = {428},
       number = {4},
        pages = {2840-2856},
          doi = {10.1093/mnras/sts210},
archivePrefix = {arXiv},
       eprint = {1206.5006},
 primaryClass = {astro-ph.IM},
       adsurl = {https://ui.adsabs.harvard.edu/abs/2013MNRAS.428.2840H}
}

@ARTICLE{ChakrabartiBlitz2009,
       author = {{Chakrabarti}, Sukanya and {Blitz}, Leo},
        title = "{Tidal imprints of a dark subhalo on the outskirts of the Milky Way}",
      journal = {\mnras},
         year = 2009,
        month = oct,
       volume = {399},
       number = {1},
        pages = {L118-L122},
          doi = {10.1111/j.1745-3933.2009.00735.x},
archivePrefix = {arXiv},
       eprint = {0812.0821},
 primaryClass = {astro-ph},
       adsurl = {https://ui.adsabs.harvard.edu/abs/2009MNRAS.399L.118C}
}

@ARTICLE{Sofue2020,
       author = {{Sofue}, Yoshiaki},
        title = "{Rotation Curve of the Milky Way and the Dark Matter Density}",
      journal = {Galaxies},
         year = 2020,
        month = apr,
       volume = {8},
       number = {2},
          eid = {37},
        pages = {37},
          doi = {10.3390/galaxies8020037},
archivePrefix = {arXiv},
       eprint = {2004.11688},
 primaryClass = {astro-ph.GA},
       adsurl = {https://ui.adsabs.harvard.edu/abs/2020Galax...8...37S}
}

@ARTICLE{Read2014,
       author = {{Read}, J.~I.},
        title = "{The local dark matter density}",
      journal = {Journal of Physics G Nuclear Physics},
         year = 2014,
        month = jun,
       volume = {41},
       number = {6},
          eid = {063101},
        pages = {063101},
          doi = {10.1088/0954-3899/41/6/063101},
archivePrefix = {arXiv},
       eprint = {1404.1938},
 primaryClass = {astro-ph.GA},
       adsurl = {https://ui.adsabs.harvard.edu/abs/2014JPhG...41f3101R}
}

@ARTICLE{PetersenPenarrubia2021,
       author = {{Petersen}, Michael S. and {Pe{\~n}arrubia}, Jorge},
        title = "{Detection of the Milky Way reflex motion due to the Large Magellanic Cloud infall}",
      journal = {Nature Astronomy},
         year = 2021,
        month = jan,
       volume = {5},
        pages = {251-255},
          doi = {10.1038/s41550-020-01254-3},
archivePrefix = {arXiv},
       eprint = {2011.10581},
 primaryClass = {astro-ph.GA},
       adsurl = {https://ui.adsabs.harvard.edu/abs/2021NatAs...5..251P}
}

@ARTICLE{Chandrasekhar1943,
       author = {{Chandrasekhar}, S.},
        title = "{Dynamical Friction. I. General Considerations: the Coefficient of Dynamical Friction.}",
      journal = {\apj},
         year = 1943,
        month = mar,
       volume = {97},
        pages = {255},
          doi = {10.1086/144517},
       adsurl = {https://ui.adsabs.harvard.edu/abs/1943ApJ....97..255C}
}

@ARTICLE{BonacaPrice-Whelan2025,
       author = {{Bonaca}, Ana and {Price-Whelan}, Adrian M.},
        title = "{Stellar streams in the Gaia era}",
      journal = {\nar},
         year = 2025,
        month = jun,
       volume = {100},
          eid = {101713},
        pages = {101713},
          doi = {10.1016/j.newar.2024.101713},
archivePrefix = {arXiv},
       eprint = {2405.19410},
 primaryClass = {astro-ph.GA},
       adsurl = {https://ui.adsabs.harvard.edu/abs/2025NewAR.10001713B}
}

@ARTICLE{Ebadi2025,
       author = {{Ebadi}, Reza and {Strokov}, Vladimir and {Tanin}, Erwin H. and {Berti}, Emanuele and {Walsworth}, Ronald L.},
        title = "{LISA double white dwarf binaries as Galactic accelerometers}",
      journal = {\prd},
         year = 2025,
        month = feb,
       volume = {111},
       number = {4},
          eid = {044023},
        pages = {044023},
          doi = {10.1103/PhysRevD.111.044023},
archivePrefix = {arXiv},
       eprint = {2405.13109},
 primaryClass = {gr-qc},
       adsurl = {https://ui.adsabs.harvard.edu/abs/2025PhRvD.111d4023E}
}

@ARTICLE{SilverwoodEasther2019,
       author = {{Silverwood}, Hamish and {Easther}, Richard},
        title = "{Stellar accelerations and the galactic gravitational field}",
      journal = {\pasa},
         year = 2019,
        month = oct,
       volume = {36},
          eid = {e038},
        pages = {e038},
          doi = {10.1017/pasa.2019.25},
archivePrefix = {arXiv},
       eprint = {1812.07581},
 primaryClass = {astro-ph.GA},
       adsurl = {https://ui.adsabs.harvard.edu/abs/2019PASA...36...38S}
}

@ARTICLE{BertoneHooper2018,
       author = {{Bertone}, Gianfranco and {Hooper}, Dan},
        title = "{History of dark matter}",
      journal = {Reviews of Modern Physics},
         year = 2018,
        month = oct,
       volume = {90},
       number = {4},
          eid = {045002},
        pages = {045002},
          doi = {10.1103/RevModPhys.90.045002},
archivePrefix = {arXiv},
       eprint = {1605.04909},
 primaryClass = {astro-ph.CO},
       adsurl = {https://ui.adsabs.harvard.edu/abs/2018RvMP...90d5002B}
}

@ARTICLE{Oort1932,
       author = {{Oort}, J.~H.},
        title = "{The force exerted by the stellar system in the direction perpendicular to the galactic plane and some related problems}",
      journal = {\bain},
         year = 1932,
        month = aug,
       volume = {6},
        pages = {249},
       adsurl = {https://ui.adsabs.harvard.edu/abs/1932BAN.....6..249O}
}

@ARTICLE{HuntVasiliev2025,
       author = {{Hunt}, Jason A.~S. and {Vasiliev}, Eugene},
        title = "{Milky Way dynamics in light of Gaia}",
      journal = {\nar},
         year = 2025,
        month = jun,
       volume = {100},
          eid = {101721},
        pages = {101721},
          doi = {10.1016/j.newar.2024.101721},
archivePrefix = {arXiv},
       eprint = {2501.04075},
 primaryClass = {astro-ph.GA},
       adsurl = {https://ui.adsabs.harvard.edu/abs/2025NewAR.10001721H}
}

@ARTICLE{Kawata2024,
       author = {{Kawata}, Daisuke and {Grand}, Robert J.~J. and {Hunt}, Jason A.~S. and {Ciuc{\u{a}}}, Ioana},
        title = "{Milky Way Disk}",
      journal = {arXiv e-prints},
         year = 2024,
        month = dec,
          eid = {arXiv:2412.12252},
        pages = {arXiv:2412.12252},
          doi = {10.48550/arXiv.2412.12252},
archivePrefix = {arXiv},
       eprint = {2412.12252},
 primaryClass = {astro-ph.GA},
       adsurl = {https://ui.adsabs.harvard.edu/abs/2024arXiv241212252K}
}

@ARTICLE{Soding2025,
       author = {{S{\"o}ding}, Laurin and {Bartel}, Ruben and {Mertsch}, Philipp},
        title = "{Local dark matter density from Gaia DR3 K-dwarfs using Gaussian processes}",
      journal = {arXiv e-prints},
         year = 2025,
        month = jun,
          eid = {arXiv:2506.02956},
        pages = {arXiv:2506.02956},
          doi = {10.48550/arXiv.2506.02956},
archivePrefix = {arXiv},
       eprint = {2506.02956},
 primaryClass = {astro-ph.GA},
       adsurl = {https://ui.adsabs.harvard.edu/abs/2025arXiv250602956S}
}

@ARTICLE{Sun2025,
       author = {{Sun}, Guang-Chen and {Wang}, Qiao and {Mao}, Shude and {Li}, Yichao and {Long}, Richard J. and {Ding}, Ping-Jie and {Wang}, Yougang and {Zhang}, Xin and {Chen}, Xuelei},
        title = "{Dynamical Models of the Milky Way in Action Space with LAMOST DR8 and Gaia EDR3}",
      journal = {\apj},
         year = 2025,
        month = mar,
       volume = {982},
       number = {1},
          eid = {37},
        pages = {37},
          doi = {10.3847/1538-4357/adb57e},
archivePrefix = {arXiv},
       eprint = {2502.08164},
 primaryClass = {astro-ph.GA},
       adsurl = {https://ui.adsabs.harvard.edu/abs/2025ApJ...982...37S}
}

@ARTICLE{Ibata2024,
       author = {{Ibata}, Rodrigo and {Malhan}, Khyati and {Tenachi}, Wassim and {Ardern-Arentsen}, Anke and {Bellazzini}, Michele and {Bianchini}, Paolo and {Bonifacio}, Piercarlo and {Caffau}, Elisabetta and {Diakogiannis}, Foivos and {Errani}, Raphael and {Famaey}, Benoit and {Ferrone}, Salvatore and {Martin}, Nicolas F. and {di Matteo}, Paola and {Monari}, Giacomo and {Renaud}, Florent and {Starkenburg}, Else and {Thomas}, Guillaume and {Viswanathan}, Akshara and {Yuan}, Zhen},
        title = "{Charting the Galactic Acceleration Field. II. A Global Mass Model of the Milky Way from the STREAMFINDER Atlas of Stellar Streams Detected in Gaia DR3}",
      journal = {\apj},
         year = 2024,
        month = jun,
       volume = {967},
       number = {2},
          eid = {89},
        pages = {89},
          doi = {10.3847/1538-4357/ad382d},
archivePrefix = {arXiv},
       eprint = {2311.17202},
 primaryClass = {astro-ph.GA},
       adsurl = {https://ui.adsabs.harvard.edu/abs/2024ApJ...967...89I}
}

@ARTICLE{Staudt2024,
       author = {{Staudt}, Patrick G. and {Bullock}, James S. and {Boylan-Kolchin}, Michael and {Kirkby}, David and {Wetzel}, Andrew and {Ou}, Xiaowei},
        title = "{Sliding into DM: determining the local dark matter density and speed distribution using only the local circular speed of the galaxy}",
      journal = {\jcap},
         year = 2024,
        month = aug,
       volume = {2024},
       number = {8},
          eid = {022},
        pages = {022},
          doi = {10.1088/1475-7516/2024/08/022},
archivePrefix = {arXiv},
       eprint = {2403.04122},
 primaryClass = {astro-ph.GA},
       adsurl = {https://ui.adsabs.harvard.edu/abs/2024JCAP...08..022S}
}

@ARTICLE{Guo2024,
       author = {{Guo}, Rui and {Li}, Zhao-Yu and {Shen}, Juntai and {Mao}, Shude and {Liu}, Chao},
        title = "{Measuring the Milky Way Vertical Potential with the Phase Snail in a Model-independent Way}",
      journal = {\apj},
         year = 2024,
        month = jan,
       volume = {960},
       number = {2},
          eid = {133},
        pages = {133},
          doi = {10.3847/1538-4357/ad037b},
archivePrefix = {arXiv},
       eprint = {2310.10225},
 primaryClass = {astro-ph.GA},
       adsurl = {https://ui.adsabs.harvard.edu/abs/2024ApJ...960..133G}
}

@ARTICLE{BinneyVasiliev2023,
       author = {{Binney}, James and {Vasiliev}, Eugene},
        title = "{Self-consistent models of our Galaxy}",
      journal = {\mnras},
         year = 2023,
        month = apr,
       volume = {520},
       number = {2},
        pages = {1832-1847},
          doi = {10.1093/mnras/stad094},
archivePrefix = {arXiv},
       eprint = {2206.03523},
 primaryClass = {astro-ph.GA},
       adsurl = {https://ui.adsabs.harvard.edu/abs/2023MNRAS.520.1832B}
}

@ARTICLE{Widmark2021,
       author = {{Widmark}, A. and {Laporte}, C.~F.~P. and {de Salas}, P.~F. and {Monari}, G.},
        title = "{Weighing the Galactic disk using phase-space spirals. II. Most stringent constraints on a thin dark disk using Gaia EDR3}",
      journal = {\aap},
         year = 2021,
        month = sep,
       volume = {653},
          eid = {A86},
        pages = {A86},
          doi = {10.1051/0004-6361/202141466},
archivePrefix = {arXiv},
       eprint = {2105.14030},
 primaryClass = {astro-ph.GA},
       adsurl = {https://ui.adsabs.harvard.edu/abs/2021A&A...653A..86W}
}

@ARTICLE{Nitschai2021,
       author = {{Nitschai}, Maria Selina and {Eilers}, Anna-Christina and {Neumayer}, Nadine and {Cappellari}, Michele and {Rix}, Hans-Walter},
        title = "{Dynamical Model of the Milky Way Using APOGEE and Gaia Data}",
      journal = {\apj},
         year = 2021,
        month = aug,
       volume = {916},
       number = {2},
          eid = {112},
        pages = {112},
          doi = {10.3847/1538-4357/ac04b5},
archivePrefix = {arXiv},
       eprint = {2106.05286},
 primaryClass = {astro-ph.GA},
       adsurl = {https://ui.adsabs.harvard.edu/abs/2021ApJ...916..112N}
}

@ARTICLE{Cautun2020,
       author = {{Cautun}, Marius and {Ben{\'\i}tez-Llambay}, Alejandro and {Deason}, Alis J. and {Frenk}, Carlos S. and {Fattahi}, Azadeh and {G{\'o}mez}, Facundo A. and {Grand}, Robert J.~J. and {Oman}, Kyle A. and {Navarro}, Julio F. and {Simpson}, Christine M.},
        title = "{The milky way total mass profile as inferred from Gaia DR2}",
      journal = {\mnras},
         year = 2020,
        month = may,
       volume = {494},
       number = {3},
        pages = {4291-4313},
          doi = {10.1093/mnras/staa1017},
archivePrefix = {arXiv},
       eprint = {1911.04557},
 primaryClass = {astro-ph.GA},
       adsurl = {https://ui.adsabs.harvard.edu/abs/2020MNRAS.494.4291C}
}

@INPROCEEDINGS{Wardana2020,
       author = {{Wardana}, M. Dafa and {Wulandari}, Hesti and {Sulistiyowati} and {Khatami}, Akbar H.},
        title = "{Determination of the local dark matter density using K-dwarfs from Gaia DR2}",
    booktitle = {European Physical Journal Web of Conferences},
         year = 2020,
       series = {European Physical Journal Web of Conferences},
       volume = {240},
        month = dec,
          eid = {04002},
        pages = {04002},
          doi = {10.1051/epjconf/202024004002},
       adsurl = {https://ui.adsabs.harvard.edu/abs/2020EPJWC.24004002W}
}

@ARTICLE{Eilers2019,
       author = {{Eilers}, Anna-Christina and {Hogg}, David W. and {Rix}, Hans-Walter and {Ness}, Melissa K.},
        title = "{The Circular Velocity Curve of the Milky Way from 5 to 25 kpc}",
      journal = {\apj},
         year = 2019,
        month = jan,
       volume = {871},
       number = {1},
          eid = {120},
        pages = {120},
          doi = {10.3847/1538-4357/aaf648},
archivePrefix = {arXiv},
       eprint = {1810.09466},
 primaryClass = {astro-ph.GA},
       adsurl = {https://ui.adsabs.harvard.edu/abs/2019ApJ...871..120E}
}

@ARTICLE{Sivertsson2018,
       author = {{Sivertsson}, S. and {Silverwood}, H. and {Read}, J.~I. and {Bertone}, G. and {Steger}, P.},
        title = "{The local dark matter density from SDSS-SEGUE G-dwarfs}",
      journal = {\mnras},
         year = 2018,
        month = aug,
       volume = {478},
       number = {2},
        pages = {1677-1693},
          doi = {10.1093/mnras/sty977},
archivePrefix = {arXiv},
       eprint = {1708.07836},
 primaryClass = {astro-ph.GA},
       adsurl = {https://ui.adsabs.harvard.edu/abs/2018MNRAS.478.1677S}
}

@ARTICLE{Xia2016,
       author = {{Xia}, Qiran and {Liu}, Chao and {Mao}, Shude and {Song}, Yingyi and {Zhang}, Lan and {Long}, R.~J. and {Zhang}, Yong and {Hou}, Yonghui and {Wang}, Yuefei and {Wu}, Yue},
        title = "{Determining the local dark matter density with LAMOST data}",
      journal = {\mnras},
         year = 2016,
        month = jun,
       volume = {458},
       number = {4},
        pages = {3839-3850},
          doi = {10.1093/mnras/stw565},
archivePrefix = {arXiv},
       eprint = {1510.06810},
 primaryClass = {astro-ph.GA},
       adsurl = {https://ui.adsabs.harvard.edu/abs/2016MNRAS.458.3839X}
}

@ARTICLE{Huang2016,
       author = {{Huang}, Y. and {Liu}, X. -W. and {Yuan}, H. -B. and {Xiang}, M. -S. and {Zhang}, H. -W. and {Chen}, B. -Q. and {Ren}, J. -J. and {Wang}, C. and {Zhang}, Y. and {Hou}, Y. -H. and {Wang}, Y. -F. and {Cao}, Z. -H.},
        title = "{The Milky Way's rotation curve out to 100 kpc and its constraint on the Galactic mass distribution}",
      journal = {\mnras},
         year = 2016,
        month = dec,
       volume = {463},
       number = {3},
        pages = {2623-2639},
          doi = {10.1093/mnras/stw2096},
archivePrefix = {arXiv},
       eprint = {1604.01216},
 primaryClass = {astro-ph.GA},
       adsurl = {https://ui.adsabs.harvard.edu/abs/2016MNRAS.463.2623H}
}

@ARTICLE{YurinSpringel2014,
       author = {{Yurin}, Denis and {Springel}, Volker},
        title = "{An iterative method for the construction of N-body galaxy models in collisionless equilibrium}",
      journal = {\mnras},
         year = 2014,
        month = oct,
       volume = {444},
       number = {1},
        pages = {62-79},
          doi = {10.1093/mnras/stu1421},
archivePrefix = {arXiv},
       eprint = {1402.1623},
 primaryClass = {astro-ph.CO},
       adsurl = {https://ui.adsabs.harvard.edu/abs/2014MNRAS.444...62Y}
}

@ARTICLE{Gadget4,
       author = {{Springel}, Volker and {Pakmor}, R{\"u}diger and {Zier}, Oliver and {Reinecke}, Martin},
        title = "{Simulating cosmic structure formation with the GADGET-4 code}",
      journal = {\mnras},
         year = 2021,
        month = sep,
       volume = {506},
       number = {2},
        pages = {2871-2949},
          doi = {10.1093/mnras/stab1855},
archivePrefix = {arXiv},
       eprint = {2010.03567},
 primaryClass = {astro-ph.IM},
       adsurl = {https://ui.adsabs.harvard.edu/abs/2021MNRAS.506.2871S}
}

@ARTICLE{HagenHelmi2018,
       author = {{Hagen}, Jorrit H.~J. and {Helmi}, Amina},
        title = "{The vertical force in the solar neighbourhood using red clump stars in TGAS and RAVE. Constraints on the local dark matter density}",
      journal = {\aap},
         year = 2018,
        month = jul,
       volume = {615},
          eid = {A99},
        pages = {A99},
          doi = {10.1051/0004-6361/201832903},
archivePrefix = {arXiv},
       eprint = {1802.09291},
 primaryClass = {astro-ph.GA},
       adsurl = {https://ui.adsabs.harvard.edu/abs/2018A&A...615A..99H}
}

@ARTICLE{Nibauer2025,
       author = {{Nibauer}, Jacob and {Bonaca}, Ana},
        title = "{Galactic Accelerations from the GD-1 Stream Suggest a Tilted Dark Matter Halo}",
      journal = {\apjl},
         year = 2025,
        month = may,
       volume = {985},
       number = {1},
          eid = {L22},
        pages = {L22},
          doi = {10.3847/2041-8213/add0a9},
archivePrefix = {arXiv},
       eprint = {2504.07187},
 primaryClass = {astro-ph.GA},
       adsurl = {https://ui.adsabs.harvard.edu/abs/2025ApJ...985L..22N}
}

@ARTICLE{Xiang2018,
       author = {{Xiang}, Maosheng and {Shi}, Jianrong and {Liu}, Xiaowei and {Yuan}, Haibo and {Chen}, Bingqiu and {Huang}, Yang and {Wang}, Chun and {Wu}, Yaqian and {Tian}, Zhijia and {Huo}, Zhiying and {Zhang}, Huawei and {Zhang}, Meng},
        title = "{Stellar Mass Distribution and Star Formation History of the Galactic Disk Revealed by Mono-age Stellar Populations from LAMOST}",
      journal = {\apjs},
         year = 2018,
        month = aug,
       volume = {237},
       number = {2},
          eid = {33},
        pages = {33},
          doi = {10.3847/1538-4365/aad237},
archivePrefix = {arXiv},
       eprint = {1807.04592},
 primaryClass = {astro-ph.GA},
       adsurl = {https://ui.adsabs.harvard.edu/abs/2018ApJS..237...33X}
}

@ARTICLE{Horta2024,
       author = {{Horta}, Danny and {Price-Whelan}, Adrian M. and {Hogg}, David W. and {Johnston}, Kathryn V. and {Widrow}, Lawrence and {Dalcanton}, Julianne J. and {Ness}, Melissa K. and {Hunt}, Jason A.~S.},
        title = "{Orbital Torus Imaging: Acceleration, Density, and Dark Matter in the Galactic Disk Measured with Element Abundance Gradients}",
      journal = {\apj},
         year = 2024,
        month = feb,
       volume = {962},
       number = {2},
          eid = {165},
        pages = {165},
          doi = {10.3847/1538-4357/ad16e8},
archivePrefix = {arXiv},
       eprint = {2312.07664},
 primaryClass = {astro-ph.GA},
       adsurl = {https://ui.adsabs.harvard.edu/abs/2024ApJ...962..165H}
}

@ARTICLE{BinneyVasiliev2024,
       author = {{Binney}, James and {Vasiliev}, Eugene},
        title = "{Chemodynamical models of our Galaxy}",
      journal = {\mnras},
         year = 2024,
        month = jan,
       volume = {527},
       number = {2},
        pages = {1915-1934},
          doi = {10.1093/mnras/stad3312},
archivePrefix = {arXiv},
       eprint = {2306.11602},
 primaryClass = {astro-ph.GA},
       adsurl = {https://ui.adsabs.harvard.edu/abs/2024MNRAS.527.1915B}
}

@ARTICLE{Bienayme2024,
       author = {{Bienaym{\'e}}, O. and {Robin}, A.~C. and {Salomon}, J. -B. and {Reyl{\'e}}, C.},
        title = "{Dark matter in the Milky Way: Measurements up to 3 kpc from the Galactic plane above the Sun}",
      journal = {\aap},
         year = 2024,
        month = sep,
       volume = {689},
          eid = {A280},
        pages = {A280},
          doi = {10.1051/0004-6361/202450327},
archivePrefix = {arXiv},
       eprint = {2406.08158},
 primaryClass = {astro-ph.GA},
       adsurl = {https://ui.adsabs.harvard.edu/abs/2024A&A...689A.280B}
}

@ARTICLE{Lim2025,
       author = {{Lim}, Sung Hak and {Putney}, Eric and {Buckley}, Matthew R. and {Shih}, David},
        title = "{Mapping dark matter in the Milky Way using normalizing flows and Gaia DR3}",
      journal = {\jcap},
         year = 2025,
        month = jan,
       volume = {2025},
       number = {1},
          eid = {021},
        pages = {021},
          doi = {10.1088/1475-7516/2025/01/021},
archivePrefix = {arXiv},
       eprint = {2305.13358},
 primaryClass = {astro-ph.GA},
       adsurl = {https://ui.adsabs.harvard.edu/abs/2025JCAP...01..021L}
}

@ARTICLE{Donlon2025,
      title={Empirical Modeling of Magnetic Braking in Millisecond Pulsars to Measure the Local Dark Matter Density and Effects of Orbiting Satellite Galaxies}, 
      author={{Donlon}, T. II and {Chakrabarti}, S. and {Widrow}, L. M. and {Vanderwaal}, S. and {Ransom}, S. and {Ramirez-Ruiz}, E.},
      year={2025},
      eprint={2501.03409},
      archivePrefix={arXiv},
      primaryClass={astro-ph.HE},
      url={https://arxiv.org/abs/2501.03409}, 
}

@ARTICLE{SivertssonRead2022,
       author = {{Sivertsson}, S. and {Read}, J.~I. and {Silverwood}, H. and {de Salas}, P.~F. and {Malhan}, K. and {Widmark}, A. and {Laporte}, C.~F.~P. and {Garbari}, S. and {Freese}, K.},
        title = "{Estimating the local dark matter density in a non-axisymmetric wobbling disc}",
      journal = {\mnras},
         year = 2022,
        month = apr,
       volume = {511},
       number = {2},
        pages = {1977-1991},
          doi = {10.1093/mnras/stac094},
archivePrefix = {arXiv},
       eprint = {2201.01822},
 primaryClass = {astro-ph.GA},
       adsurl = {https://ui.adsabs.harvard.edu/abs/2022MNRAS.511.1977S}
}

@ARTICLE{Banik2017,
       author = {{Banik}, Nilanjan and {Widrow}, Lawrence M. and {Dodelson}, Scott},
        title = "{Galactoseismology and the local density of dark matter}",
      journal = {\mnras},
         year = 2017,
        month = feb,
       volume = {464},
       number = {4},
        pages = {3775-3783},
          doi = {10.1093/mnras/stw2603},
archivePrefix = {arXiv},
       eprint = {1608.03338},
 primaryClass = {astro-ph.GA},
       adsurl = {https://ui.adsabs.harvard.edu/abs/2017MNRAS.464.3775B}
}

@ARTICLE{BovyTremaine2012,
       author = {{Bovy}, Jo and {Tremaine}, Scott},
        title = "{On the Local Dark Matter Density}",
      journal = {\apj},
         year = 2012,
        month = sep,
       volume = {756},
       number = {1},
          eid = {89},
        pages = {89},
          doi = {10.1088/0004-637X/756/1/89},
archivePrefix = {arXiv},
       eprint = {1205.4033},
 primaryClass = {astro-ph.GA},
       adsurl = {https://ui.adsabs.harvard.edu/abs/2012ApJ...756...89B}
}

@ARTICLE{Zhang2013,
       author = {{Zhang}, Lan and {Rix}, Hans-Walter and {van de Ven}, Glenn and {Bovy}, Jo and {Liu}, Chao and {Zhao}, Gang},
        title = "{The Gravitational Potential near the Sun from SEGUE K-dwarf Kinematics}",
      journal = {\apj},
         year = 2013,
        month = aug,
       volume = {772},
       number = {2},
          eid = {108},
        pages = {108},
          doi = {10.1088/0004-637X/772/2/108},
archivePrefix = {arXiv},
       eprint = {1209.0256},
 primaryClass = {astro-ph.GA},
       adsurl = {https://ui.adsabs.harvard.edu/abs/2013ApJ...772..108Z}
}

@ARTICLE{Garbari2012,
       author = {{Garbari}, Silvia and {Liu}, Chao and {Read}, Justin I. and {Lake}, George},
        title = "{A new determination of the local dark matter density from the kinematics of K dwarfs}",
      journal = {\mnras},
         year = 2012,
        month = sep,
       volume = {425},
       number = {2},
        pages = {1445-1458},
          doi = {10.1111/j.1365-2966.2012.21608.x},
archivePrefix = {arXiv},
       eprint = {1206.0015},
 primaryClass = {astro-ph.GA},
       adsurl = {https://ui.adsabs.harvard.edu/abs/2012MNRAS.425.1445G}
}

@ARTICLE{KuijkenGilmore1991,
       author = {{Kuijken}, Konrad and {Gilmore}, Gerard},
        title = "{The Galactic Disk Surface Mass Density and the Galactic Force K Z at Z = 1.1 Kiloparsecs}",
      journal = {\apjl},
         year = 1991,
        month = jan,
       volume = {367},
        pages = {L9},
          doi = {10.1086/185920},
       adsurl = {https://ui.adsabs.harvard.edu/abs/1991ApJ...367L...9K}
}

@ARTICLE{Cheng2024,
       author = {{Cheng}, Xinlun and {Anguiano}, Borja and {Majewski}, Steven R. and {Arras}, Phil},
        title = "{The surface mass density of the Milky Way: does the traditional K$_{Z}$ approach work in the context of new surveys?}",
      journal = {\mnras},
         year = 2024,
        month = jan,
       volume = {527},
       number = {1},
        pages = {959-976},
          doi = {10.1093/mnras/stad3013},
archivePrefix = {arXiv},
       eprint = {2309.17405},
 primaryClass = {astro-ph.GA},
       adsurl = {https://ui.adsabs.harvard.edu/abs/2024MNRAS.527..959C}
}

@ARTICLE{HainesDonghia2019,
       author = {{Haines}, Tim and {D'Onghia}, Elena and {Famaey}, Benoit and {Laporte}, Chervin and {Hernquist}, Lars},
        title = "{Implications of a Time-varying Galactic Potential for Determinations of the Dynamical Surface Density}",
      journal = {\apjl},
         year = 2019,
        month = jul,
       volume = {879},
       number = {1},
          eid = {L15},
        pages = {L15},
          doi = {10.3847/2041-8213/ab25f3},
archivePrefix = {arXiv},
       eprint = {1903.00607},
 primaryClass = {astro-ph.GA},
       adsurl = {https://ui.adsabs.harvard.edu/abs/2019ApJ...879L..15H}
}

@ARTICLE{Vasiliev2021,
       author = {{Vasiliev}, Eugene and {Belokurov}, Vasily and {Erkal}, Denis},
        title = "{Tango for three: Sagittarius, LMC, and the Milky Way}",
      journal = {\mnras},
         year = 2021,
        month = feb,
       volume = {501},
       number = {2},
        pages = {2279-2304},
          doi = {10.1093/mnras/staa3673},
archivePrefix = {arXiv},
       eprint = {2009.10726},
 primaryClass = {astro-ph.GA},
       adsurl = {https://ui.adsabs.harvard.edu/abs/2021MNRAS.501.2279V}
}

@ARTICLE{Donlon2024,
       author = {{Donlon}, Thomas, II and {Chakrabarti}, Sukanya and {Widrow}, Lawrence M. and {Lam}, Michael T. and {Chang}, Philip and {Quillen}, Alice C.},
        title = "{Galactic Structure From Binary Pulsar Accelerations: Beyond Smooth Models}",
      journal = {arXiv e-prints},
         year = 2024,
        month = jan,
          eid = {arXiv:2401.15808},
        pages = {arXiv:2401.15808},
          doi = {10.48550/arXiv.2401.15808},
archivePrefix = {arXiv},
       eprint = {2401.15808},
 primaryClass = {astro-ph.GA},
       adsurl = {https://ui.adsabs.harvard.edu/abs/2024arXiv240115808D}
}

@ARTICLE{Widrow2012,
       author = {{Widrow}, Lawrence M. and {Gardner}, Susan and {Yanny}, Brian and {Dodelson}, Scott and {Chen}, Hsin-Yu},
        title = "{Galactoseismology: Discovery of Vertical Waves in the Galactic Disk}",
      journal = {\apjl},
         year = 2012,
        month = may,
       volume = {750},
       number = {2},
          eid = {L41},
        pages = {L41},
          doi = {10.1088/2041-8205/750/2/L41},
archivePrefix = {arXiv},
       eprint = {1203.6861},
 primaryClass = {astro-ph.GA},
       adsurl = {https://ui.adsabs.harvard.edu/abs/2012ApJ...750L..41W}
}

@BOOK{BinneyTremaine2008,
       author = {{Binney}, James and {Tremaine}, Scott},
        title = "{Galactic Dynamics: Second Edition}",
         year = 2008,
       adsurl = {https://ui.adsabs.harvard.edu/abs/2008gady.book.....B}
}

@ARTICLE{NFW,
       author = {{Navarro}, Julio F. and {Frenk}, Carlos S. and {White}, Simon D.~M.},
        title = "{A Universal Density Profile from Hierarchical Clustering}",
      journal = {\apj},
         year = 1997,
        month = dec,
       volume = {490},
       number = {2},
        pages = {493-508},
          doi = {10.1086/304888},
archivePrefix = {arXiv},
       eprint = {astro-ph/9611107},
 primaryClass = {astro-ph},
       adsurl = {https://ui.adsabs.harvard.edu/abs/1997ApJ...490..493N}
}

@ARTICLE{Hernquist1990,
       author = {{Hernquist}, Lars},
        title = "{An Analytical Model for Spherical Galaxies and Bulges}",
      journal = {\apj},
         year = 1990,
        month = jun,
       volume = {356},
        pages = {359},
          doi = {10.1086/168845},
       adsurl = {https://ui.adsabs.harvard.edu/abs/1990ApJ...356..359H}
}

@ARTICLE{Chakrabarti2022,
       author = {{Chakrabarti}, Sukanya and {Stevens}, Daniel J. and {Wright}, Jason and {Rafikov}, Roman R. and {Chang}, Philip and {Beatty}, Thomas and {Huber}, Daniel},
        title = "{Eclipse Timing the Milky Way's Gravitational Potential}",
      journal = {\apjl},
         year = 2022,
        month = apr,
       volume = {928},
       number = {2},
          eid = {L17},
        pages = {L17},
          doi = {10.3847/2041-8213/ac5c43},
archivePrefix = {arXiv},
       eprint = {2112.08231},
 primaryClass = {astro-ph.GA},
       adsurl = {https://ui.adsabs.harvard.edu/abs/2022ApJ...928L..17C}
}

@ARTICLE{Buch2019,
       author = {{Buch}, Jatan and {Leung}, John Shing Chau and {Fan}, JiJi},
        title = "{Using Gaia DR2 to constrain local dark matter density and thin dark disk}",
      journal = {\jcap},
         year = 2019,
        month = apr,
       volume = {2019},
       number = {4},
          eid = {026},
        pages = {026},
          doi = {10.1088/1475-7516/2019/04/026},
archivePrefix = {arXiv},
       eprint = {1808.05603},
 primaryClass = {astro-ph.GA},
       adsurl = {https://ui.adsabs.harvard.edu/abs/2019JCAP...04..026B}
}

@ARTICLE{Guo2020,
       author = {{Guo}, Rui and {Liu}, Chao and {Mao}, Shude and {Xue}, Xiang-Xiang and {Long}, R.~J. and {Zhang}, Lan},
        title = "{Measuring the local dark matter density with LAMOST DR5 and Gaia DR2}",
      journal = {\mnras},
         year = 2020,
        month = jul,
       volume = {495},
       number = {4},
        pages = {4828-4844},
          doi = {10.1093/mnras/staa1483},
archivePrefix = {arXiv},
       eprint = {2005.12018},
 primaryClass = {astro-ph.GA},
       adsurl = {https://ui.adsabs.harvard.edu/abs/2020MNRAS.495.4828G}
}

@ARTICLE{McKee2015,
       author = {{McKee}, Christopher F. and {Parravano}, Antonio and {Hollenbach}, David J.},
        title = "{Stars, Gas, and Dark Matter in the Solar Neighborhood}",
      journal = {\apj},
         year = 2015,
        month = nov,
       volume = {814},
       number = {1},
          eid = {13},
        pages = {13},
          doi = {10.1088/0004-637X/814/1/13},
archivePrefix = {arXiv},
       eprint = {1509.05334},
 primaryClass = {astro-ph.GA},
       adsurl = {https://ui.adsabs.harvard.edu/abs/2015ApJ...814...13M}
}

@ARTICLE{BlandHawthornGerhard2016,
       author = {{Bland-Hawthorn}, Joss and {Gerhard}, Ortwin},
        title = "{The Galaxy in Context: Structural, Kinematic, and Integrated Properties}",
      journal = {\araa},
         year = 2016,
        month = sep,
       volume = {54},
        pages = {529-596},
          doi = {10.1146/annurev-astro-081915-023441},
archivePrefix = {arXiv},
       eprint = {1602.07702},
 primaryClass = {astro-ph.GA},
       adsurl = {https://ui.adsabs.harvard.edu/abs/2016ARA&A..54..529B}
}

@ARTICLE{McMillan2017,
       author = {{McMillan}, Paul J.},
        title = "{The mass distribution and gravitational potential of the Milky Way}",
      journal = {\mnras},
         year = 2017,
        month = feb,
       volume = {465},
       number = {1},
        pages = {76-94},
          doi = {10.1093/mnras/stw2759},
archivePrefix = {arXiv},
       eprint = {1608.00971},
 primaryClass = {astro-ph.GA},
       adsurl = {https://ui.adsabs.harvard.edu/abs/2017MNRAS.465...76M}
}

@ARTICLE{Chakrabarti2020,
       author = {{Chakrabarti}, Sukanya and {Wright}, Jason and {Chang}, Philip and {Quillen}, Alice and {Craig}, Peter and {Territo}, Joey and {D'Onghia}, Elena and {Johnston}, Kathryn V. and {De Rosa}, Robert J. and {Huber}, Daniel and {Rhode}, Katherine L. and {Nielsen}, Eric},
        title = "{Toward a Direct Measure of the Galactic Acceleration}",
      journal = {\apjl},
         year = 2020,
        month = oct,
       volume = {902},
       number = {1},
          eid = {L28},
        pages = {L28},
          doi = {10.3847/2041-8213/abb9b5},
archivePrefix = {arXiv},
       eprint = {2007.15097},
 primaryClass = {astro-ph.GA},
       adsurl = {https://ui.adsabs.harvard.edu/abs/2020ApJ...902L..28C}
}

@ARTICLE{Chakrabarti2021,
       author = {{Chakrabarti}, Sukanya and {Chang}, Philip and {Lam}, Michael T. and {Vigeland}, Sarah J. and {Quillen}, Alice C.},
        title = "{A Measurement of the Galactic Plane Mass Density from Binary Pulsar Accelerations}",
      journal = {\apjl},
         year = 2021,
        month = feb,
       volume = {907},
       number = {2},
          eid = {L26},
        pages = {L26},
          doi = {10.3847/2041-8213/abd635},
archivePrefix = {arXiv},
       eprint = {2010.04018},
 primaryClass = {astro-ph.GA},
       adsurl = {https://ui.adsabs.harvard.edu/abs/2021ApJ...907L..26C}
}

@ARTICLE{Xu2015,
       author = {{Xu}, Yan and {Newberg}, Heidi Jo and {Carlin}, Jeffrey L. and {Liu}, Chao and {Deng}, Licai and {Li}, Jing and {Sch{\"o}nrich}, Ralph and {Yanny}, Brian},
        title = "{Rings and Radial Waves in the Disk of the Milky Way}",
      journal = {\apj},
         year = 2015,
        month = mar,
       volume = {801},
       number = {2},
          eid = {105},
        pages = {105},
          doi = {10.1088/0004-637X/801/2/105},
archivePrefix = {arXiv},
       eprint = {1503.00257},
 primaryClass = {astro-ph.GA},
       adsurl = {https://ui.adsabs.harvard.edu/abs/2015ApJ...801..105X}
}

@ARTICLE{Antoja2018,
       author = {{Antoja}, T. and {Helmi}, A. and {Romero-G{\'o}mez}, M. and {Katz}, D. and {Babusiaux}, C. and {Drimmel}, R. and {Evans}, D.~W. and {Figueras}, F. and {Poggio}, E. and {Reyl{\'e}}, C. and {Robin}, A.~C. and {Seabroke}, G. and {Soubiran}, C.},
        title = "{A dynamically young and perturbed Milky Way disk}",
      journal = {\nat},
         year = 2018,
        month = sep,
       volume = {561},
       number = {7723},
        pages = {360-362},
          doi = {10.1038/s41586-018-0510-7},
archivePrefix = {arXiv},
       eprint = {1804.10196},
 primaryClass = {astro-ph.GA},
       adsurl = {https://ui.adsabs.harvard.edu/abs/2018Natur.561..360A}
}

@ARTICLE{Arora2022,
       author = {{Arora}, Arpit and {Sanderson}, Robyn E. and {Panithanpaisal}, Nondh and {Cunningham}, Emily C. and {Wetzel}, Andrew and {Garavito-Camargo}, Nicol{\'a}s},
        title = "{On the Stability of Tidal Streams in Action Space}",
      journal = {\apj},
         year = 2022,
        month = nov,
       volume = {939},
       number = {1},
          eid = {2},
        pages = {2},
          doi = {10.3847/1538-4357/ac93fb},
archivePrefix = {arXiv},
       eprint = {2207.13481},
 primaryClass = {astro-ph.GA},
       adsurl = {https://ui.adsabs.harvard.edu/abs/2022ApJ...939....2A}
}

%
%
%

\begin{table*}[]
    \centering
    \begin{tabular}{|lllr|} \hline \hline
        $\rho_{0,\mathrm{DM}}$ & No. Sources & Type of Study & Reference \\
        (M$_\odot$/pc$^{3}$) & & & \\ \hline 
        0.0084 $\pm$ 0.0005 & 21,700 & Rotation Curve & \cite{Huang2016} \\
        0.018 $\pm$ 0.0054 & 1427 & Jeans Modeling & \cite{Xia2016} \\
        0.0106 $\pm$ 0.0053 & -- & Synthesis & \cite{McMillan2017} \\
        0.012 $\pm$ 0.002 & 13,000 & Jeans Modeling & \cite{Sivertsson2018} \\
        0.018 $\pm$ 0.002 & 26,653 & Jeans Modeling & \cite{HagenHelmi2018} \\
        0.016 $\pm$ 0.010 & 51,374 & DF Fitting & \cite{Buch2019} \\
        0.0079 $\pm$ 0.00079 & 23,129 & Rotation Curve & \cite{Eilers2019} \\
        0.0133$^{+0.0024}_{-0.0022}$ & 90,000 & Jeans Modeling & \cite{Guo2020} \\
        0.0116 $\pm$ 0.0012 & 95,543 & Jeans Modeling & \cite{Wardana2020} \\
        0.0088 $\pm$ 0.0005 & 23,000 & Rotation Curve & \cite{Cautun2020} \\
        0.00892 $\pm$ 0.00056 & 2,857,689 & Synthesis & \cite{Nitschai2021} \\
        0.0085 $\pm$ 0.0039 & 812,250 & Phase Space Spiral & \cite{Widmark2021} \\
        -0.004$^{+0.05}_{-0.02}$ & 14 & Pulsar Accelerations & \cite{Chakrabarti2021} \\
        0.0121 & 1,000,000 & DF Fitting & \cite{BinneyVasiliev2023} \\
        0.0150 $\pm$ 0.0031 & 33,800,000 & Phase Space Spiral & \cite{Guo2024} \\
        0.011 & -- & DF Fitting & \cite{BinneyVasiliev2024} \\
        0.011 $\pm$ 0.002 & -- & Rotation Curve & \cite{Staudt2024} \\
        0.0128 $\pm$ 0.0008 & 45,000$^1$ & DF Fitting & \cite{Bienayme2024} \\
        0.0114 $\pm$ 0.0007 & 33,000,000 & Tidal Streams & \cite{Ibata2024} \\
        0.011 $\pm$ 0.002 & 94,685 & DF Fitting & \cite{Horta2024} \\
        -0.01 $\pm$ 0.018 & 26 & Pulsar Accelerations & \cite{Donlon2024} \\
        0.012 $\pm$ 0.001 & 5,811,956 & DF Fitting & \cite{Lim2025} \\
        0.0115 $\pm$ 0.00115 & 86,109 & DF Fitting & \cite{Sun2025} \\
        0.0131 $\pm$ 0.0041 & 200,000 & Jeans Modeling & \cite{Soding2025} \\
        0.0098 $\pm$ 0.0025 & 53 & Pulsar Accelerations & \cite{Donlon2025} \\        
        \hline \hline
    \end{tabular}
    \caption{Selected measurements of the dark matter density in the Galactic midplane from the last decade. The data are plotted in Figure \ref{fig:rho_dm}. 
    $^1$Estimated from their Figure 5.}
    \label{tab:dm_density}
\end{table*}

\appendix 

\section{References for Figure 1} \label{sec:app_fig1_references}

The references for Figure \ref{fig:rho_dm} are provided in Table \ref{tab:dm_density}.

\section{Error in Jeans Modeling Due to Variance in the Slope of the Rotation Curve} \label{app:jeans_error_rotcurve}

\begin{figure*}
    \centering
    \includegraphics[width=\linewidth]{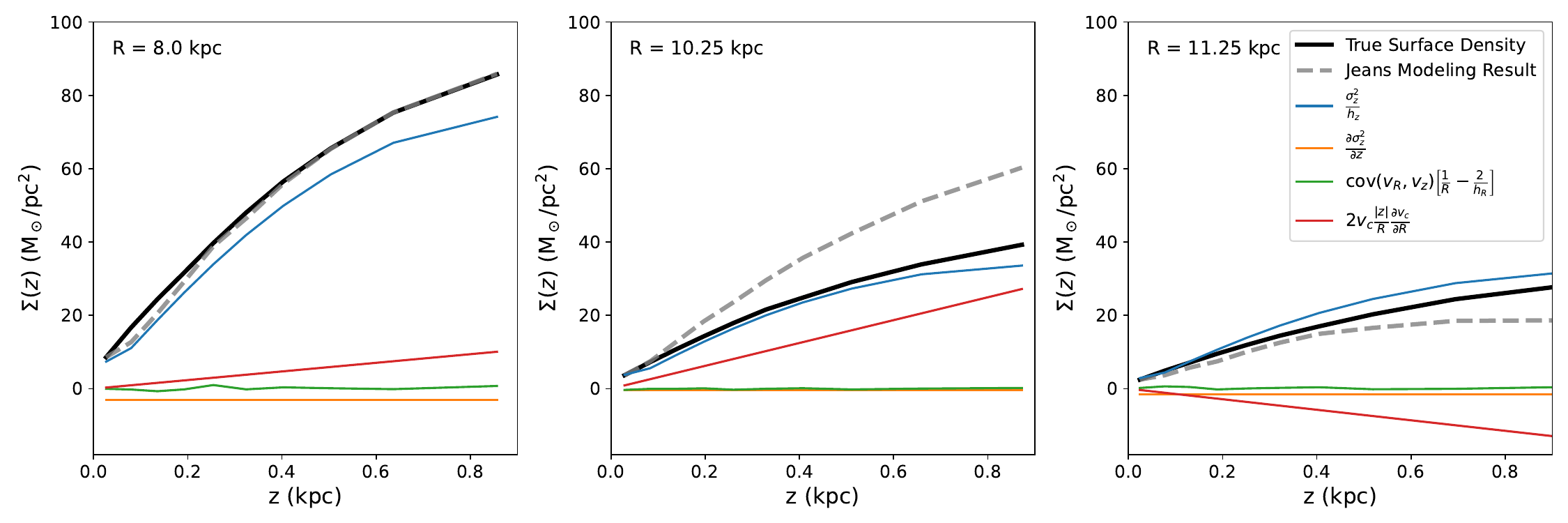}
    \caption{Individual contributions of each term from Equation \ref{eq:jeans} for the simulation data at different distances. The final term, which is a correction based on the slope of the rotation curve, is over-/under-estimated when using a local slope in the rotation curve (shown here). These issues disappear if a smooth approximation to the global rotation curve is used instead. Each contribution term was multiplied by a factor of $1/(2\pi G)$ to obtain the correct units.}
    \label{fig_app:jeans_contributions}
\end{figure*}

Equation \ref{eq:jeans} states how to compute surface density from the phase-space distribution of stars. The final term on the right-hand side of this equation is a first-order correction which accounts for a change in the slope of the rotation curve across the data. The size of this correction term is sizable compared to the second and third terms, and can substantially change the inferred surface density if it is incorrect or ignored. 

We plot the contribution of each term on the overall inferred surface density at a range of heights for simulated data in Figure \ref{fig_app:jeans_contributions}. Each panel shows the data for a cylinder with a radius of 1 kpc located at a different distance. In the first panel, the inferred value of the surface density from Jeans modeling agrees with the true surface density computed from the simulation data. In the other panels, the inferred surface density is too high and too low, respectively. The over/under-estimate of the surface density is correlated with the rotation curve correction, suggesting that this correction may be the source of this issue. 

We find that these issues disappear when we instead use a smooth polynomial fit to the overall rotation curve, rather than the local slope of the rotation curve at each point. The rotation curve often contains dynamical features that cause this correction term to be inaccurate; this is unrelated to the true dynamics of the stars in that volume, which will pass through such a region in a relatively small amount of time without being strongly perturbed. In contrast, a smooth fit to the global rotation curve produces the shape of the rotation curve that is actually felt over long timescales by disk stars, and will influence their motion.

\section{Effect of disequilibrium terms on direct acceleration techniques} \label{app:diseq_on_dams}

\begin{figure}
    \centering
    \includegraphics[width=\linewidth]{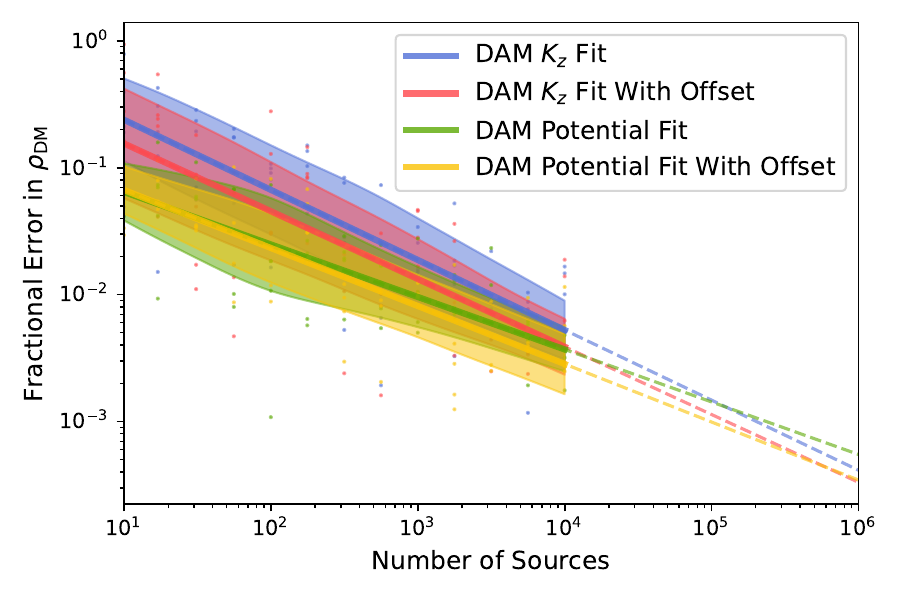}
    \caption{Comparison of the direct acceleration fitting methods, with and without an additional disequilibrium term. Adding an additional term to the analytical model for the disk acceleration profile slightly improves the performance of the model. However, adding a vertical displacement between the disk and halo potential components did not significantly improve the method's ability to recover the local density of dark matter compared to an equilibrium potential model. }
    \label{fig_app:unc_diseq_offsets}
\end{figure}

The procedures for computing the local dark matter density in Section \ref{sec:direct_acc_modeling} and \ref{sec:precision_per_source} are time-independent models, in that they do not explicitly take into account any departure from equilibrium in the distribution function. We show that this is not necessarily an issue because these methods are resilient to disequilibrium in a way that Jeans modeling is not. However, one can explicitly add additional structure to these methods that incorporate disequilibrium features, in an attempt to better model the local dark matter distribution. 

The first method involves fitting an acceleration profile to the observed acceleration data. In this case, it has the form of a single $\sech^2$ disk, which is a good approximation of the vertical structure of the disks in the simulations. One can account for disequilibrium structure by adding additional terms to the acceleration profile. For example, \begin{align}
    a_z(z) &= A h_z \tanh\left(\frac{z}{h_z}\right) + \sum_{i=0} a^{(i)}_{z,0} z^i \\ \nonumber
    &= A h_z \tanh\left(\frac{z}{h_z}\right) + a_{z,0} + a'_{z,0}z + ...
\end{align} illustrates adding a power series of the acceleration field to the $\sech^2$ disk model. Such features are realistically plausible; if the disk near the Sun is vertically displaced relative to the rest of the Galaxy, a bulk overall acceleration would be apparent across the direct acceleration measurements as a nonzero value of $a_{z,0}$. 

\begin{figure*}
    \centering
    \includegraphics[width=\linewidth]{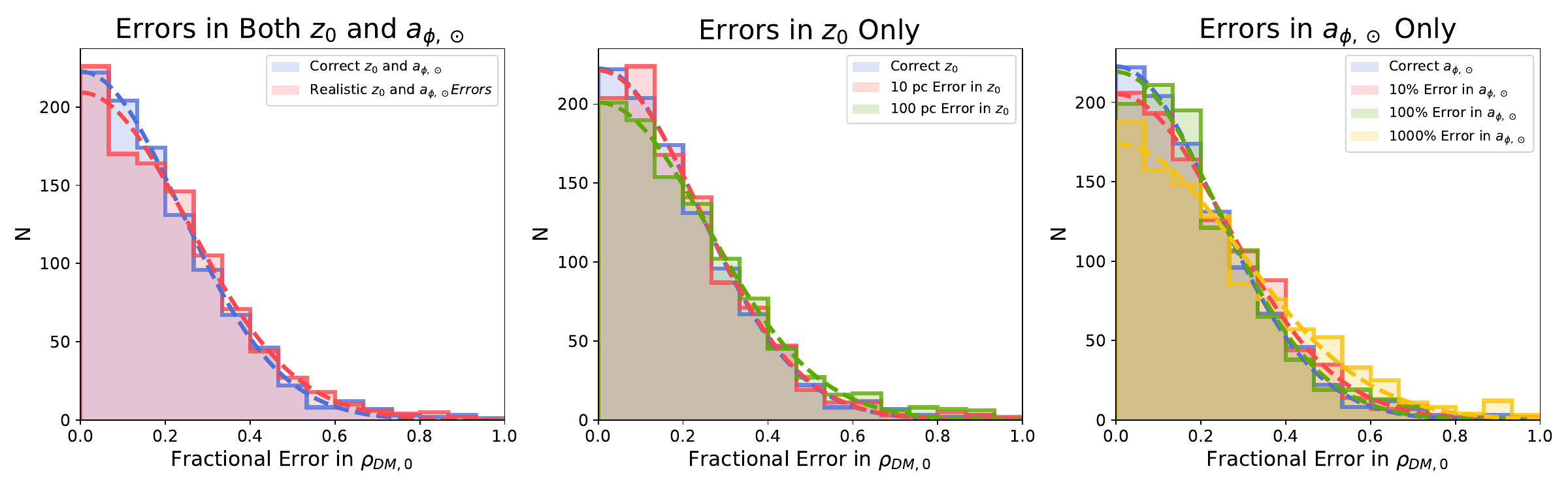}
    \caption{The effect of errors in the observed distance of the Sun from the Galactic midplane ($z_0$) and the Sun's azimuthal acceleration ($a_\phi$) on the ability to determine the local midplane DM density from direct acceleration measurements. The error in $\rho_{DM,0}$ remains essentially unchanged even for uncertainties in $a_0$ and $a_\phi$ that are much larger than the observational constraints, indicating that the direct acceleration method is resilient to observational uncertainties in these values.}
    \label{fig:app_z0_aphi0_errs}
\end{figure*}

The second method incorporates the procedure of \cite{Donlon2025} by allowing the center of mass of the disk and halo components of the potential to be vertically displaced from one another. This sort of feature is expected to occur due to the passage of nearby dwarf galaxies, such as the LMC. 

We re-ran the procedure from the main text on these two adjusted methods, which are plotted in Figure \ref{fig_app:unc_diseq_offsets}. Adding an additional term to the $a_z$ model in the first method slightly improves the performance of the model. However, adding a vertical displacement between the disk and halo potential components did not significantly improve the method's ability to recover the local density of dark matter compared to an equilibrium potential model. Because these adjustments do not significantly improve the performance of the models, we conclude that it is acceptable to ignore additional disequilibrium terms for the purposes of this work.

\section{Error in the Sun's Vertical Position and Deviation from Axisymmetry} \label{app:z0_aphi_err}

The direct acceleration analysis above makes the assumption that we know the Sun's vertical position exactly. This is not the case; the Sun's distance from the Galactic midplane ($z_0$) is only know at a $\sim$5--10 pc level \citep{BlandHawthornGerhard2016}. Similarly, small deviations from axisymmetry could take the form of a several-percent deviation in the Sun's azimuthal acceleration $a_\phi$ (this type of acceleration could also be caused by local disruptions to the acceleration field near the Sun, such as a nearby unknown celestial body or structure in the local distribution of dark matter). 

Here, we test the effect of realistic uncertainties on these quantities on the determination of the local dark matter density. We randomly place 50 direct acceleration measurements in a 1 kpc box around the Sun, assign each point its acceleration from the isolated simulation, and then use the ``$K_z$ fit'' method to obtain an estimate for $\rho_{DM,0}.$ This is then compared to the true midplane DM density in the simulation. Then, the local DM density was determined in the case of: \begin{enumerate}
    \item Realistic errors in both $z_0$ (10 pc) and $a_\phi$ (10\%), 
    \item No error in $a_\phi$, but varying the error in $z_0$ between 10 pc and 100 pc, and
    \item No error in $z_0$, but varying the error in $a_\phi$ between 10\% and 1000\% of the correct value.
\end{enumerate} This procedure was repeated over 1000 random draws; the results are shown in Figure \ref{fig:app_z0_aphi0_errs}. 

Overall, the accuracy in the determination of $\rho_{DM,0}$ was essentially unchanged for all of these cases. Even increasing the uncertainties in $z_0$ and $a_\phi$ well beyond their observational constraints only resulted in a small loss of precision in $\rho_{DM,0}.$ Based on this, we can conclude that the direct acceleration method is very resilient to uncertainties in these quantities. 

\end{document}